\documentclass[a4paper,11pt]{article}
\pdfoutput=1 

\usepackage{jcappub}
\usepackage{gensymb}
\usepackage{float}
\usepackage{rotating}

\title{Detection prospects for heavy WIMP dark matter around M31* in microwave band}

\author{Andrei E. Egorov}
\affiliation{Institute of Physics, University of Belgrade, Pregrevica 118, 11080 Belgrade, Serbia}

\emailAdd{aegorov@runbox.com}

\abstract{This work analyzes the detection prospects for weakly interacting massive particles (WIMPs) in dark matter (DM) density spike around the supermassive black hole (SMBH) in Andromeda galaxy M31 in microwave band. WIMP annihilation produces relativistic electrons and positrons, which generate the synchrotron emission in strong magnetic field of the accretion flow medium at microwave frequencies. Properties of this emission, as well as their dependence on DM and medium parameters, were modeled. Atacama Large Millimeter Array (ALMA) sensitivity to this emission and underlying WIMP parameters was estimated. It was obtained that M31* in microwave band represents the unique target, where DM spike signal can be detected in case of realistic spike and halo density profiles. In such case ALMA may reach the thermal s-wave annihilating WIMPs with masses up to around 1 TeV with just 10--100 hours exposure. Optimistic parameter configurations may allow probing all possible WIMP masses up to 100 TeV. Related systematic uncertainties are discussed.}

\begin{document}
\maketitle
\flushbottom

\section{\label{sec:i}Introduction and motivation}

DM phenomenon was discovered already almost a century ago \cite{1937ApJ....86..217Z,1936ApJ....83...23S}. However, it still presents a puzzle in our knowledge about Universe -- exact physical nature of DM persists to stay unknown despite many decades of active research. A very wide variety of DM candidates has been considered -- one can see e.g. the contemporary review \cite{SciPostPhysRev.1}. Historically WIMPs represent the most probable and motivated candidate, which represents hypothetical supersymmetric particles \cite{1984NuPhB.238..453E}. One of the main WIMP search strategies is indirect or astrophysical; which is based on the anticipated possibility of WIMP self-annihilation with production of energetic Standard Model particles, which in turn would manifest themselves in various astrophysical objects. WIMPs are indeed a very wide class of particles \cite{2025EPJC...85..152A}. Considering even the simple basic thermally produced s-wave annihilating WIMPs, all indirect detection techniques (including gamma-ray observations of Milky Way satellites, measurements of cosmic-ray antiprotons, radio observations of Local Group galaxies and others) have probed reliably so far only relatively light WIMPs with the mass $m_x \lesssim$ 0.1 TeV \cite{SciPostPhysRev.1}. This is conditioned by fast decrease of the intensity of all annihilation signals with WIMP mass increase. However, theoretically WIMPs may have the mass up to so-called unitarity limit $\approx$ 100 TeV \cite{2019PhRvD.100d3029S}! Thus, the major part of anticipated mass range has not been probed so far despite of extensive multi-messenger searches during decades. And a lot of further development is required in order to achieve either discovery or full exclusion of at least these WIMPs of observationally-favorable type. The specific example of such "good" and heavy WIMP is Higgsino with $m_x \approx 1.1$ TeV \cite{2023PhRvL.130t1001D}.

The required progress can be realized by combining of both: increase of detectors' sensitivity, and selection of unique targets with very high DM density and very faint interfering astrophysical emissions. DM-generated emission intensity is defined by the annihilation rate; which is basically proportional to $\langle\sigma v\rangle \rho^2/m_x^2$, where $\langle\sigma v\rangle$ is velocity-averaged product of the annihilation cross section and particle relative velocity, $\rho$ is DM density. Therefore, in order to compensate the intensity fading at high WIMP masses we need to look at the densest DM targets. In practice all these conditions can be fulfilled for TeV-scale s-wave annihilating WIMPs in, presumably, just two settings: observations of the Galactic center region in very-high energy band and multiwavelength observations of density spikes around nearby SMBHs. The first strategy will be implemented by Cherenkov Telescope Array (CTA)\footnote{\url{https://www.ctao.org}} and Southern Wide-field Gamma-ray Observatories (SWGO)\footnote{\url{https://www.swgo.org}}. Their sensitivity predictions can be summarized as follows \cite{2021JCAP...01..057A,2019JCAP...12..061V}. In a good case scenario of cuspy DM density profile in Milky Way (MW), these telescopes are expected to probe the thermal WIMPs in mass range $\sim$(0.1--10) TeV. However, if the density profile has a substantial core with radius $\sim$ 1 kpc, \emph{any} WIMPs would \emph{not} be reachable! This uncertainty is quite fundamental, because DM density can not be determined with satisfactory precision in inner
few kiloparsecs of galaxies like MW. Thus, even in the optimistic case CTA together with preceding facilities (e.g. Fermi-LAT) will reach $m_x \sim 10$ TeV, leaving the heaviest WIMPs unprobed. Moreover, significant uncertainties related to a separation of various emission components may degrade this picture.

In this situation DM density spikes around SMBHs seem to be unique targets, which may enable -- at least, potentially -- the full test of thermal WIMPs. This approach employs the possibility of very strong DM concentration by SMBH gravity in its local vicinity of influence. This idea was elaborated for the first time already a while ago by \cite{1999PhRvL..83.1719G}. Since that many works have been accomplished on this topic: \cite{2014PhRvL.113o1302F,2015PhRvD..92d3510L,2015PhRvL.115w1302S,2020PhRvD.102b3030C,2023PhRvD.108j3042C,2023JCAP...08..063B,2025PhRvD.111k5033P,2025EPJC...85.1100K,2026arXiv260223348C,2026JHEP...02..050W} and others. However, despite all these extensive efforts, a complete exploration of this type of targets seems to not have been achieved yet. The main subject of analysis was prompt gamma-ray emission due to DM annihilation in MW* and M87*. M31* was not studied at all, although it is a massive and nearby SMBH with very faint observed emission. The latter circumstance may form some bias toward lowered interest. However, for the purpose of WIMP searches, fainter emission (in relevant wavebands) implies better sensitivity. These considerations motivated me to compare quantitatively without a bias all nearby SMBHs and identify the most promising targets among them. This was done in the frame of my preceding work \cite{2026JCAP...04..002E}. The conducted analysis revealed, that in fact only MW* and M31* may produce any detectable flux of gamma rays under realistic model assumptions. Both targets provide comparable sensitivity to DM signal: MW* is expected to be slightly brighter, but M31* has weaker confusing astrophysical emissions. Moreover, according to my CTA sensitivity estimates, M31* can deliver even much stronger WIMP constraints than MW* for certain observational scenarios. And M87*, which received a bit of attention in the past, appeared to be one of the faintest objects.

As known, DM annihilation generates secondary emissions besides prompt gamma rays. Thus, relativistic electrons and positrons ($e^\pm$) produced by annihilating WIMPs should emit synchrotron in a strong magnetic field (MF) in SMBH environment. And it is very interesting to evaluate WIMP detection prospects concerning such synchrotron emission in M31*, which was already proved to be very promising target in gamma rays. This evaluation constitutes the subject of my current work, which naturally extends the previous project \cite{2026JCAP...04..002E} to a full multiwavelength study of M31*.

From observational point of view, there is also a good motivation to consider the synchrotron emission, which typically falls into radio and microwave bands. On one hand, synchrotron is more model-dependent in comparison with the prompt gamma, since MF knowledge is needed. However, this disadvantage could be compensated by much better angular resolution and overall sensitivity of radio-microwave telescopes. A bright example is very famous imaging of SMBHs by the Event Horizon Telescope (EHT)\footnote{\url{https://eventhorizontelescope.org}} Collaboration. Considering ALMA\footnote{\url{https://almaobservatory.org}}, it is able to reach $\sim\mu$Jy sensitivity levels and subarcsecond angular resolutions, which may provide very good opportunities for M31* DM spike exploration.

Here I employ majorly M31* DM spike model developed in \cite{2026JCAP...04..002E}, use the same notation and do not rewrite all the similar model equations for brevity. The synchrotron emission is defined by MF distribution inside the spike. A comprehensive model of the accretion flow around SMBH has to be constructed in order to specify MF distribution. Hence, the system being modeled here is more complicated and has two components -- DM spike and accretion flow. Thus, the population of highly-energetic $e^\pm$ from DM resides inside accretion flow; which contains MF, thermal population of electrons and ions, radiation field. I made the following basic assumptions: spherical symmetry of the whole system, absence of relativistic effects in SMBH close vicinity, absence of effects related to SMBH rotation and to p-wave contribution into DM annihilation cross section. These assumptions were discussed in \cite[section 1.2]{2026JCAP...04..002E}. I worked in the frame of ultrarelativistic limit for DM $e^\pm$, i.e. the energy-momentum relation for them is $E=pc$. I also made the synchrotron emission estimates for MW* for self-check and comparison.

The next subsection briefly reviews the main published studies, which were dedicated to the synchrotron emission from DM density spikes. Then the content is organized as follows: section \ref{sec:gen} describes all key steps of the emission flux computation, section \ref{sec:flow} explains the model of accretion flow medium and its thermal emission, section \ref{sec:tr} describes the solution of transport equation for $e^\pm$ produced by DM annihilation, section \ref{sec:obt} analyzes the spectral and morphological properties of the obtained emission, section \ref{sec:res} presents the main results -- ALMA sensitivity to WIMP parameters, and section \ref{sec:con} discusses the results and proposes further plans.

\subsection{\label{ssec:i}Brief review of previous works on the subject}

The first mature work on DM constraints from the synchrotron emission due to WIMP annihilation in the density spike was probably \cite{2004JCAP...05..007A}. Those authors excluded the possibility of very steep (adiabatic) spike density profile using radio observations of MW* and a modest theoretical framework. Later \cite{2008PhRvD..78d3505R} refined significantly and extended MW* framework to a fully-multiwavelength analysis. This work deduced an absence of any WIMP signatures, derived meaningful constraints and predicted a quite limited potential of future gamma-ray telescopes. 

Then \cite{2015PhRvD..92d3510L} employed M87* for the first time and stated quite strong WIMP constraints from multiwavelength analysis of this object. However, their constraints are unrealistically strong, likely, because when the authors modeled DM density spike, they set 20 kpc for M87 DM halo scale radius, i.e. the value from MW. However, M87 DM halo differs very much from that in MW -- the latter is lighter by about 2 (!) orders of magnitude (\cite{2025PhRvD.111k5033P,2026JCAP...04..002E} contains detailed discussions on this weakness). Hence, such extrapolation is incorrect, overestimates the spike density and, therefore, respective constraints. Soon after the same group published \cite{2017PhRvD..96f3008L}, where the very-high angular resolution imaging of M87* DM spike by EHT was studied. However, the same mistake with the central halo density was repeated. Also, the authors focused mainly on the densest possible (adiabatic) spike, which is a low-probable scenario, as discussed below in the next section. And very recently \cite{2025PhRvL.135l1001C} conducted the similar study of EHT sensitivity to WIMP parameters in M87*. Overall their physical model has a deep level of elaboration. However, the authors seem to copy the outlined problems from \cite{2015PhRvD..92d3510L,2017PhRvD..96f3008L} without any critical reassessment.

Summarizing, we may note that the accomplished works had certain bias toward an observationally favorable DM spike version rather than physically realistic one. And M31* has not been studied yet. My work aims to progress on both of these gaps. 

\section{\label{sec:gen}General formalism for emission flux computation}

This section describes the general algorithm of computation of flux density $F(\nu)$ of the synchrotron emission from DM density spike:
\begin{equation}\label{eq:F}
F(\nu) = \frac{1}{4\pi d^2}\iiint j(\nu,R)d^3R = \frac{2}{d^2}\int\limits_{3R_\bullet}^{R_m}\int\limits_{E_0}^{m_xc^2} \psi(R,E)P_e(\nu,R,E)R^2dEdR,
\end{equation}
where spherical symmetry and medium transparency are assumed, $d$ is the distance to SMBH from observer, $R$ is the radial distance from SMBH, $j(\nu,R)$ is the volumic emissivity of $e^\pm$ produced by DM, $\psi(R,E)$ is the equilibrium spectral concentration of \emph{one} specie of $e^\pm$ (both of them create additional prefactor 2) and $P_e(\nu,R,E)$ is the spectral synchrotron emission power of one particle. Let us discuss the integration limits. Integration over space covers the spike region, which we want to collect emission from. The inner spike boundary is typically assumed to be at two Schwarzschild radii (2$R_\bullet,~R_\bullet=2GM/c^2$), which is the radius of last stable orbit for particles \cite{2013PhRvD..88f3522S}. However, I intentionally ignored everything beneath 3$R_\bullet$, which is an approximate border of development of strong-gravity effects. Particularly, the photon gravitational redshift exceeds $\approx 10\%$, if a photon is emitted below 3$R_\bullet$ in radial direction. Thus, in order to avoid flux overestimation in the frame of my weak-gravity approximation, I set this inner boundary, though the innermost spike region inside few $R_\bullet$ contributes a very little flux fraction anyway. And the upper bound $R_m$ can be arbitrary up to the spike radius. Integration over energy goes from the smallest relevant value, which was empirically found to be $E_0 = 10~\text{MeV}\approx 20m_ec^2$ and satisfies ultrarelativistic approximation.

The synchrotron spectral power $P_e(\nu,R,E)$ was computed through the standard way taking into account averaging over random angles between MF and $e^\pm$ momenta directions \cite{1988A&A...196..327C,2011ApJ...737...21L}:
\begin{equation}\label{eq:P}
	\begin{split}
P_e(\nu,R,E) = \frac{2\pi m_e^2c^3e^2\nu}{\sqrt{3}E^2}\int\limits_0^\pi d\theta \sin\theta \int\limits_{x/\sin\theta}^\infty dyK_{5/3}(y) &= \\
=\frac{2\pi^2m_e^2c^3e^2\nu}{\sqrt{3}E^2}(W_{0,\frac{4}{3}}(x)W_{0,\frac{1}{3}}(x)-W_{\frac{1}{2},\frac{5}{6}}(x)W_{-\frac{1}{2},\frac{5}{6}}(x)) &\approx \frac{2\pi^2m_e^2c^3e^2\nu}{\sqrt{3}E^2(0.869x^{2/3}+x\exp(x))}, \\
x &= 4\pi m_e^3c^5\nu/(3eE^2B(R)),
	\end{split}
\end{equation}
where $W(x)$ are so-called Whittaker functions and $B(R)$ is MF strength. The exact expression of double integral above through Whittaker functions was originally derived in \cite{1986A&A...164L..16C}, is very useful and simplifies computations significantly. However, Whittaker functions are still quite heavy in a massive numerical evaluation. That is why the second approximation from \cite{2011ApJ...737...21L} was employed, which gives the precision not worse than 20\% and is sufficient for our model.

The spectral concentration $\psi(R,E)$ comes from the solution of $e^\pm$ stationary transport equation, which has the following form in its general version written for the phase-space distribution function $f(R,\vec{p})$ (e.g. \cite[eq. (S1)]{2025PhRvL.135l1001C}, but certain signs can be wrong there):
\begin{equation}\label{eq:f}
\nabla(\vec{V}f)+\nabla_{\vec{p}}(\dot{\vec{p}}f)-\nabla(D\nabla f)-\nabla_{\vec{p}}(D_{pp}\nabla_{\vec{p}}f) = q_p(R,p) = \frac{\langle\sigma v\rangle \rho^2(R)}{2m_x^2}\frac{dN_e}{d^3p}(p,m_x),	
\end{equation}
where $\vec{V}(R)$ is the velocity field of accretion flow, $\dot{\vec{p}}=\dot{\vec{p}}_r+\dot{\vec{p}}_a$ is $e^\pm$ momenta change due to energy losses (mainly radiative) and adiabatic gain, $D$ is the diffusion coefficient, $D_{pp}$ describes diffusive reacceleration due to MHD turbulence, $q_p(R,p)$ is the source function in momentum representation and $dN_e/d^3p$ is the average DM $e^\pm$ spectra at production/injection from one WIMP annihilation event. Here we impose the non-trivial assumption of full coupling between DM $e^\pm$ population and the ambient plasma of accretion flow through MF. I.e., saying qualitatively, SMBH pulls the thermal plasma by gravity, the former moves with the frozen MF, which in turn drags $e^\pm$. This assumption is justified by a sufficiently slow spatial diffusion, as will be shown below.

We should model both the accretion flow medium and DM source. The first includes the spatial distributions of MF, velocity, density, temperature etc. and is described in section \ref{sec:flow} below. Here let us specify the model of DM source. $e^\pm$ spectra at production were taken from the conventional database PPPC4DMID \cite{PPPC,2011JCAP...03..051C,2011JCAP...03..019C} (they are provided in the form $dN_e/dE(E,m_x)$). Indeed they depend on both WIMP mass and primary annihilation products (channel). I obtained all the main results for two traditional representative channels $\chi\chi\rightarrow b\bar{b},\tau^+\tau^-$; although other leptonic channels, namely $e^+e^-$ and $\mu^+\mu^-$, can produce the synchrotron emission fluxes outside the range confined by $b\bar{b},\tau^+\tau^-$ cases at some WIMP masses. Also recently more refined particle source spectra databases were built, particularly CosmiXs \cite{2024JCAP...03..035A}. However, their spectra differ by not more than $\approx 20\%$ from those in PPPC4DMID for the involved primary and final annihilation products (\cite{2024JCAP...03..035A} and private communication with the authors). Such difference is quite negligible in comparison with systematic model uncertainties. 

\subsection{\label{ssec:spike}DM spike density profile}
The biggest uncertainty in modeling of the anticipated DM signal comes from the uncertainty in the spike density profile. Traditionally it is approximated by the power law $\rho(R)\propto R^{-\gamma}$. Large enough values of $\langle\sigma v\rangle$ can limit the density growth in the inner spike region due to too fast annihilation rate. This flattens the profile slope down to the well-predicted value $\gamma_{in} = 0.5$ (\cite{2016PhRvD..93l3510S}, see also \cite[eqs. (2.2)--(2.5)]{2026JCAP...04..002E} and explanation there). But the main slope value $\gamma$ is harder to predict. At beginning of the topic the authors \cite{1999PhRvL..83.1719G} considered the idealistic case of adiabatic SMBH growth in DM-only environment. In this case the spike develops very steep profile with $\gamma = 2.25\div2.5$ depending on the external halo density profile. However, later more realistic two-component SMBH environment was introduced to the model: i.e. DM and stars, which are present in any galactic nucleus. Moving stars heat and stir DM spike, which strongly depletes its density and lowers the slope down to $\gamma$ = 1.5, as was derived originally in \cite{2004PhRvL..93f1302G,2004PhRvL..92t1304M}. Many subsequent theoretical studies confirmed independently this result. Thus, \cite{2022PhRvD.106d3018S} obtained the same $\gamma$ = 1.5 throughout two independent formalisms, \cite{2026arXiv260328866S,2026arXiv260501023H} generalized the model by elaboration of evolution with redshift, addition of stellar mass function etc. Such picture diminished an initial optimism, because only dense spikes with $\gamma\gtrsim 2$ were typically found to produce a substantial detectable signal. However, very recently the authors \cite{2026arXiv260613761K} brought attention to the important detail: known members of nuclear star cluster in MW do not go below the radial distances $R\sim (10^{-4}-10^{-3})$ pc and leave DM spike majorly undisturbed/adiabatic inside this sphere. This work also found out the spike resilience to disturbances from stellar black holes, which inspiral and merge to SMBH. Hence, the spike structure can be more tricky than the long-assumed single power-law profile (with possible core due to DM annihilation) and have two zones instead: outer stirred zone with depleted density and inner preserved zone with (semi-)adiabatic profile. Such spike would be definitely brighter than that with $\gamma$ = 1.5, which revives an interest. Also looking broader, the mentioned spike models often focused on the particular case of MW or even lighter SMBHs. Heavier SMBHs in other galaxies may have different stellar environments, spike evolution paths and resulting profiles. Thus, \cite{2026arXiv260501023H} obtained $\gamma$ = 1.5 as the minimal asymptotic value, but it experiences significant variation over redshift, distance from SMBH, stellar population properties etc. And all these theoretical models have not been tested seriously by observations yet due to a complexity of such task (\cite{2024MNRAS.527.3196S,2026PhRvD.113d3052S} and discussion in \cite[section 2.1]{2026JCAP...04..002E}).

Based on all the described studies, I chose $\gamma = 1.5$ as the robust minimal (MIN) value for the model. $\gamma = 1.75$ (so-called Bahcall-Wolf cusp) was set as the medium (MED) value: it approximates effectively the mixed two-zone stirred/undisturbed spike discussed above and is considered to be a realistic value. And $\gamma = 2.0$ was chosen to be maximal (MAX) possible, optimistic value; which corresponds to a highly preserved semi-adiabatic spike. I discarded from consideration the pure adiabatic case $\gamma\approx 2.3$ mainly due to its very low probability. Also, as shown below, such dense spike would be very bright and, therefore, detected already or soon even by relatively shallow M31* observations in the microwave band, at least in the case of s-wave thermal WIMPs.

Another significant uncertainty comes from that in DM halo density value $\rho_0$ around the spike. This value serves as the basis for spike profile and can not be determined with satisfactory accuracy. The spike density profile joins continuously the external halo profile at the spike outer edge, i.e. $\rho_0=\rho(R_{sp})=\rho_\text{ext}(R_{sp})$. I accepted here consistently the same MIN--MAX M31 inner halo density profiles as in my preceding works \cite{2022PhRvD.106b3023E,2026JCAP...04..002E}. Table \ref{tab:par} below shows the respective $\rho_0$ values, as well as all other model parameter values.

And, finally, we need to specify the spike radius $R_{sp}$. It corresponds approximately to the radius of sphere of SMBH gravitational influence: $R_{sp}=bGM/\sigma_c^2$, where $\sigma_c$ is 1D (i.e. radial) velocity dispersion of the material (DM and stars) in the galactic central region and can be measured observationally. The prefactor $b\sim 1$ encompasses fine physical effects. As was pointed out in \cite[section 2.1]{2026JCAP...04..002E}, several past studies \cite{2014PhRvL.113o1302F,2015PhRvL.115w1302S,2020PhRvD.102b3030C,2023JCAP...08..063B} likely underestimated the spike radius putting $b=0.2$, while the detailed observational study of MW nuclear star cluster \cite{2018A&A...609A..27S} obtained $b\approx 0.6$ at least. At the same time, recent theoretical spike analysis \cite{2022PhRvD.106d3018S} found out $b\simeq 1$ (figures 1 and 3 there). These studies pinned down $b$ to a relatively certain value. And I set the middle value $b=0.8$ for all MIN--MAX parameter configurations. Mild uncertainty of this value plays a minor role in comparison with other model systematics.

\begin{sidewaystable}
\centering
\begin{tabular}{|c|cccc|c|}
\hline
\rule{0pt}{1em}
\textbf{Parameter} & \textbf{M31* MIN} & \textbf{M31* MED} & \textbf{M31* MAX} & \textbf{M31* ref.} & \textbf{MW*} \\
\hline
 & \multicolumn{5}{|c|}{DM density spike} \\
\hline
Distance $d$ [kpc] & \multicolumn{3}{c}{760} & {\cite{2021ApJ...920...84L}} & 8.3 {\cite{2022A&A...657L..12G}} \\
SMBH mass $M~[M_\odot]$ & $1.1\cdot 10^8$ & $1.4\cdot 10^8$ & $2.3\cdot 10^8$ & {\cite{2005ApJ...631..280B}} & $4.3\cdot 10^6$ {\cite{2022A&A...657L..12G}} \\
Central velocity dispersion $\sigma_c$ [km/s] & 170 & 150 & 130 & {\cite{2018MNRAS.481.3210B,2025MNRAS.542..669G}} & 130 {\cite{2018A&A...616A..83V}} \\
Spike radius $R_{sp}$ [pc; $10^6R_\bullet$] & 13; 1.2 & 21; 1.6 & 47; 2.1 & section \ref{ssec:spike} & 0.87; 2.1 \\
Spike angular radius $R_{sp}/d$ [arcsec] & 3.5 & 5.8 & 13 & -- & 22 \\
DM halo density at spike edge $\rho_0$ [GeV/cm$^3$] & 5.8 & 40 & 130 & {\cite{2022PhRvD.106b3023E}} & 100 {\cite{2021JCAP...01..057A}} \\
The main density profile slope $\gamma$ & 1.5 & 1.75 & 2.0 & section \ref{ssec:spike} & same as for M31* \\
\hline
 & \multicolumn{5}{|c|}{Accretion flow} \\
\hline
Viscosity parameter $\alpha$ & 0.01 & 0.0055 & 0.001 & {\cite{2012ApJ...761..129Y,2012ApJ...761..130Y,2014ARA&A..52..529Y}} & 0.0055 \\
Pressure ratio $\beta\equiv\text{\textit\textpeso}_\text{gas}/(\text{\textit\textpeso}_\text{gas}+\text{\textit\textpeso}_\text{MF})$ & 0.90 & 0.93 & 0.98 & {\cite{2012ApJ...761..129Y,2012ApJ...761..130Y,2014ARA&A..52..529Y}} & 0.93 \\
Density profile slope $\delta(R>10R_\bullet)$ & 0.85 & 0.75 & 0.65 & {\cite{2012ApJ...761..129Y,2012ApJ...761..130Y,2014ARA&A..52..529Y}} & 0.75 \\
Accretion flow radius $R_{fl}$ [pc; $10^6R_\bullet$] & 12; 1.2 & 15; 1.2 & 25; 1.2 & {\cite{2010ApJ...710..755G}} & 0.12; 0.3 {\cite{2003ApJ...591..891B}} \\
Accretion rate $\dot{M}(R_{fl})~[10^{-5}M_\odot/\text{year};~10^{-5}\dot{M}_\text{Edd}]$ & 5.8; 2.4 & 9.3; 3.0 & 25; 4.9 & {\cite{2010ApJ...710..755G}} & 0.2; 2 {\cite{2003ApJ...591..891B}}\\
MF strength $B(R=3R_\bullet)$ [G] & 5.4 & 3.4 & 2.4 & section \ref{sec:flow} & 27 \\
Electron concentration $n_e(R=3R_\bullet)$ [cm$^{-3}$] & $1.3\cdot10^5$ & $7.5\cdot10^4$ & $1.3\cdot10^5$ & section \ref{sec:flow} & $4.7\cdot10^6$ \\
Electron temperature $kT_e(R=3R_\bullet)$ [MeV] & $\lesssim2.3$ & $\lesssim2.8$ & $\lesssim2.7$ & section \ref{sec:flow} & 8.3 \\
Radiation energy density $U(R=3R_\bullet)$ [erg/cm$^3$] & \multicolumn{3}{c}{0.002} & section \ref{sec:flow} & 0.2 \\
Thermal optical depth $\tau_m(\nu=100$ GHz) & $\lesssim6\cdot10^{-4}$ & $\lesssim10^{-3}$ & $\lesssim6\cdot10^{-5}$ & section \ref{sec:flow} & 6.0 \\
\hline
\end{tabular}
\caption{\label{tab:par}All the main model parameter values for both MW* and M31* in three employed MIN--MAX parameter configurations with corresponding information sources (for MW* all parameters except $\gamma$ were fixed). $\delta(R\leqslant10R_\bullet)=1.5$ for all cases.}
\end{sidewaystable}

\section{\label{sec:flow}Accretion flow model}

Both MW* and M31* have quite similar and low normalized accretion rates, and belong to the class of radiatively inefficient advection-dominated accretion flows (ADAFs) \cite{2017ApJ...845..140Y,2025A&A...693A..24M}. The theory of such flows was developed well already a while ago (e.g. \cite{1995ApJ...452..710N}). Here I fully adopted the generic ADAF model presented in \cite{2012ApJ...761..129Y,2012ApJ...761..130Y,2014ARA&A..52..529Y}. A general feature of ADAF is geometrically quite thick accretion disk: its vertical scale height is about half of the radial-in-plane distance from center. This enables not-so-large system anisotropy and, hence, certain eligibility for our spherical symmetry approximation. Thus I work here in terms of angle-averaged physical quantities. Their radial distributions are provided by \cite[eqs. (18)]{2012ApJ...761..129Y}:
\begin{align}
\label{eq:B}
B(R) &= 6.5\cdot10^8(1-\beta)^\frac{1}{2}\alpha^{-\frac{1}{2}}\left(\frac{M}{M_\odot}\right)^{-\frac{1}{2}}\left(\frac{\dot{M}(R_{fl})}{\dot{M}_\text{Edd}}\right)^\frac{1}{2}\left(\frac{R_{fl}}{R_\bullet}\right)^{\frac{\delta}{2}-\frac{3}{4}}\left(\frac{R}{R_\bullet}\right)^{-\frac{\delta}{2}-\frac{1}{2}}~\text{G},  \\
\label{eq:ne}
n_e(R) &= 6.3\cdot10^{19}\alpha^{-1}\left(\frac{M}{M_\odot}\right)^{-1} \left(\frac{\dot{M}(R_{fl})}{\dot{M}_\text{Edd}}\right)\left(\frac{R_{fl}}{R_\bullet}\right)^{\delta-\frac{3}{2}}\left(\frac{R}{R_\bullet}\right)^{-\delta}~\text{cm}^{-3},    \\
\label{eq:V}
V_R(R) &= -1.1\cdot10^{10}\alpha(R/R_\bullet)^{-1/2}~\text{cm/s},     \\ 
\label{eq:cs}
c_s^2(R) &=	1.4\cdot10^{20}R_\bullet/R~(\text{cm/s})^2,
\end{align}
where $\beta\equiv\text{\textit\textpeso}_\text{gas}/(\text{\textit\textpeso}_\text{gas}+\text{\textit\textpeso}_\text{MF})$ is the thermal gas partial pressure, $\alpha$ is the dimensionless Shakura-Sunyaev viscosity parameter \cite{1973A&A....24..337S}, $R_{fl}$ is the accretion flow outer radius, $\dot{M}_\text{Edd} \equiv 10L_\text{Edd}/c^2 = 15Gc^3m_pm_e^2M/e^4$ is Eddington accretion rate, $n_e(R)$ is the concentration of thermal electrons, $V_R(R)$ is the flow radial speed\footnote{Note: the expression for $V_R(R)$ in \cite{2012ApJ...761..129Y} has the error -- the prefactor must be $-1.1\cdot10^{10}$, as it is in \cite{2014ARA&A..52..529Y}!} and $c_s^2\equiv(\text{\textit\textpeso}_\text{gas}+\text{\textit\textpeso}_\text{MF})/\rho$ is isothermal sound speed.

$\alpha,~\beta$  and $\delta$ parameter values were obtained in the cited accretion flow simulations and are written out in table \ref{tab:par}. It is important that these values can not be arbitrary permutated among MIN--MAX configurations. $\beta$ has a strong angular dependence: the jet region inside accretion flow has much higher degree of magnetization than the disk region. I estimated the angle-averaged $\beta$ values using \cite[figure 4]{2014ARA&A..52..529Y} and also the correlation $\alpha\beta/(1-\beta)\approx0.5$ mentioned there. The flow radius $R_{fl}$ and mass accretion rate there $\dot{M}(R_{fl})$ can be estimated as Bondi radius and rate \cite{1952MNRAS.112..195B}:
\begin{equation}\label{eq:Bondi}
R_{fl}=2GMc_{s0}^{-2},~\dot{M}(R_{fl})\approx\pi m_p\mu n_{e0}(GM)^2c_{s0}^{-3},~c_{s0}=\sqrt{5/3kT_0/(\mu m_p)},
\end{equation}
where index 0 refers to the external medium around the accretion flow. I substituted the following values for M31 according to \cite{2010ApJ...710..755G}: $\mu=0.7,~n_{e0}=0.1$ cm$^{-3},~kT_0=0.34$ keV. The resulting quantities are displayed in table \ref{tab:par}.

The gas temperature distribution is also needed for calculations of thermal emission and absorption. The temperature is obtained from the sound speed employing the gas equation of state. The gas is typically treated as ideal two-temperature plasma, which implies $\text{\textit\textpeso}_\text{gas}=n_ekT_e+n_ikT_i=kT_e(n_e+\eta n_i)$, where $\eta\equiv T_i/T_e$ is highly uncertain ion-to-electron temperature ratio. Taking into account $n_e=\rho_\text{gas}/(\mu m_p)$ and substituting the equation of state into the sound speed definition, the following expression can be derived:
\begin{equation}\label{eq:T}
\Theta_e(R)\equiv kT_e(R) = \xi(\beta,\eta,\mu)c_s^2(R)m_p,	
\end{equation}
where $\xi(\beta,\eta,\mu)\sim0.1$ encompasses fine gas properties, including helium-to-hydrogen ratio. The determination of $\eta$ value is explained below in section \ref{ssec:th}.

Another relevant model ingredient is the distribution of thermal radiation energy density -- it enters DM $e^\pm$ radiative energy losses. The thermal photon field contains, in general, two components. The main one is produced by accretion process itself. The second component is a large-scale "sea" of bulge starlight, which DM spike is immersed in. Roughly we can parametrize the radiation field density as follows (e.g. \cite[eq. (B1)]{1995ApJ...452..710N}):
\begin{equation}\label{eq:U}
U(R) \approx L/(3\pi cR^2)+U_\star,
\end{equation}
where $L$ is the bolometric flow luminosity ($L_\text{MW*}\approx10^{36}$ erg/s, $L_\text{M31*}\approx10^{37}$ erg/s  \cite{2025A&A...693A..24M}), $U_\star$ is the bulge starlight density taken to be uniform ($U_{\star\text{MW}}\approx10$ eV/cm$^3$, $U_{\star\text{M31}}\approx20$ eV/cm$^3$ \cite[section IIID]{2022PhRvD.106b3023E}).

Table \ref{tab:par} displays MF, $n_e,~\Theta_e$ and $U$ values at $R=3R_\bullet$, which is the inner edge of emission integration region in my model. The outlined theoretical model of accretion flow has to undergo a reasonable verification before application to the model of DM spike emission. A basic check can be made by comparison of my generic MW* flow medium model with that in very in-depth study by EHT collaboration \cite{2024ApJ...964L..26E}. This work obtained the following MF strength values in MW* vicinity: $B(2R_\bullet)=67^{+8}_{-9},~B(3.65R_\bullet)=26^{+3}_{-4}$ G, while the above model predicts 45 and 21 G respectively, which is a satisfactory agreement within $\approx2\sigma_s$ range. The electron concentration and temperature also agree reasonably with those shown in \cite[figure 2]{2024ApJ...964L..26E}. M31* does not have currently such data on medium properties. However, we should check at least whether M31* parameter distributions match the respective large-scale bulge values at the outer edge of accretion flow. Thus eqs. \eqref{eq:B}--\eqref{eq:U} give for M31* MED configuration at $R=R_{fl}$: $B\approx30~\mu$G, $n_e\sim1$ cm$^{-3}$, $\Theta_e\lesssim0.04$ keV, $U\approx U_\star=20$ eV/cm$^3$. MF value is in good agreement with that outlined in \cite[eqs. (6)--(8)]{2022PhRvD.106b3023E} for the bulge region. $n_e$ and $\Theta_e$ values are slightly off those obtained in \cite{2010ApJ...710..755G}. However, they are still sane, and also they are less important for DM implications.

For more advanced validation, I compare the predicted thermal synchrotron emission spectrum with the observed one. But analysis of both emissions -- the thermal and due to WIMPs -- requires knowledge of optimal spectral window, i.e. the frequency range without absorption, where we can work in the frame of transparent medium approximation, since a presence of emission absorption would complicate our task too much.

\subsection{\label{ssec:abs}Emission absorption}
In general, the dense plasma in accretion flow can absorb the microwave emission of interest by the following mechanisms: Razin effect, bremsstrahlung (free-free), thermal and self-synchrotron. Below I discuss every mechanism separately.

\textbf{Razin effect} was found out for the first time probably by \cite{Razin}. This effect is not an absorption in direct meaning, but rather a tricky effect of synchrotron radiation damping by ambient plasma due to its refraction index $\tilde{n}\neq1$. This refraction index enters the general equation for emission power \cite[eqs. (4.22)-(4.23)]{1966SvPhU...8..674G}). For the case of cold and isotropic medium $\tilde{n}^2=1-\nu_{pl}^2/\nu^2=1-e^2n_e/(\pi m_e\nu^2)$, as well-known. However, the accretion flow plasma is neither cold nor anisotropic, strictly speaking; i.e. electrons can be relativistic and the medium can be magnetized. These circumstances complicate greatly an accurate calculation of $\tilde{n}$ (e.g. \cite{Melrose}). I made the simple and basic verification of effect irrelevance using the above formula for $\tilde{n}$. Thus, the concentration $n_e\lesssim10^5$ cm$^{-3}$ from table \ref{tab:par} yields $\nu_{pl}\lesssim0.003~\text{GHz}\ll\nu\gtrsim100$ GHz and $1-\tilde{n}^2\lesssim10^{-9}$. Relativistic corrections may modify this value. However, such modifications likely would reduce it even further by (roughly) Lorentz-factor of plasma electrons \cite{Melrose}. Therefore, Razin effect should not create any attenuation of emission intensity.

\textbf{Bremsstrahlung absorption} coefficient $\kappa_b(\nu)$ is usually obtained from Kirchhoff's law assuming local thermal equilibrium: 
\begin{equation}\label{eq:jb}
j_b(\nu)=\kappa_b(\nu)I_{bb}(\nu,\Theta_e), ~ I_{bb}(\nu,\Theta_e)=2h\nu^3c^{-2}(\exp(h\nu/\Theta_e)-1)^{-1},
\end{equation}
where $I_{bb}$ is the black-body intensity. The following equation was used to calculate the total bremsstrahlung emission coefficient \cite[eq. (C7)]{2020ApJ...898...50Y}:
\begin{equation}
j_b(\nu) = \frac{8e^6}{3m_e^2 c^4}\sqrt{\frac{2\pi}{3}}\Theta_e^{-1/2}n_e^2\exp(h\nu/\Theta_e)1.2(1+2.61\Theta_e). 
\end{equation}
The corresponding optical depth can be estimated using the case of radial photon propagation:
\begin{equation}
\tau_b(\nu) \approx \int\limits_{3R_\bullet}^{R_{sp}}\kappa_b(\nu,R)dR = \int\limits_{3R_\bullet}^{R_{sp}}j_b(\nu,R)/I_{bb}(\nu,R)dR.
\end{equation}
This yields essentially zero absorption at the level $\tau_b\lesssim10^{-10}$ for M31*. Thus, bremsstrahlung mechanism is not relevant around SMBH in microwave band.

\textbf{Synchrotron absorption} is dominated by the thermal component and can be calculated by similar algorithm as above, just substituting the following emission coefficient for magnetobremsstrahlung from \cite[eq. (72)]{2011ApJ...737...21L}:
\begin{equation}\label{eq:jm}
j_m(\nu) = \frac{\sqrt{2}\pi e^2\nu_s n_e}{3cK_2(1/\Theta_e)}\left(\left(\frac{\nu}{\nu_s}\right)^\frac{1}{2}+2^\frac{11}{12}\left(\frac{\nu}{\nu_s}\right)^\frac{1}{6}\right)^2\exp\left(-\left(\frac{\nu}{\nu_s}\right)^\frac{1}{3}\right),~\nu_s=\frac{eB\Theta_e^2\sin\theta}{9\pi m_e c}.
\end{equation}
This coefficient depends on the angle $\theta$ between MF and electron velocity directions. For this reason, I included the angle-averaging into calculation of the magnetobremsstrahlung optical depth:
\begin{equation}\label{eq:tau}
\tau_m(\nu) = 0.5\int\limits_{3R_\bullet}^{R_{sp}}\int\limits_{0}^{\pi}j_m(\nu,\theta,R)\sin\theta/I_{bb}(\nu,R)dRd\theta.
\end{equation}
The estimated $\tau_m$ values is presented in table \ref{tab:par} for $\nu = 100$ GHz. And here we see a striking advantage of M31* with respect to MW*: the former's accretion flow is nearly transparent at this frequency, while MW* environment blocks all the emission from inner spike layers completely! Thus, M31* develops significant absorption at much lower frequencies: $\tau_m(\nu\approx40\text{ GHz})\approx0.1$. Regarding the synchrotron self-absorption of the emission from DM $e^\pm$, I verified its absence using the standard equation for the self-absorption coefficient (e.g., \cite[eq. (11)]{1988A&A...196..327C}\footnote{Note: the same equation in \cite[eq. (4.17)]{1966SvPhU...8..674G} misses the minus sign!}). Hence, we may conclude that only the thermal magnetobremsstrahlung absorption is relevant in the system.

\subsection{\label{ssec:th}Obtained thermal emission from accretion flow}
\begin{figure}[t]
	\centering
	\includegraphics[width=0.504\textwidth]{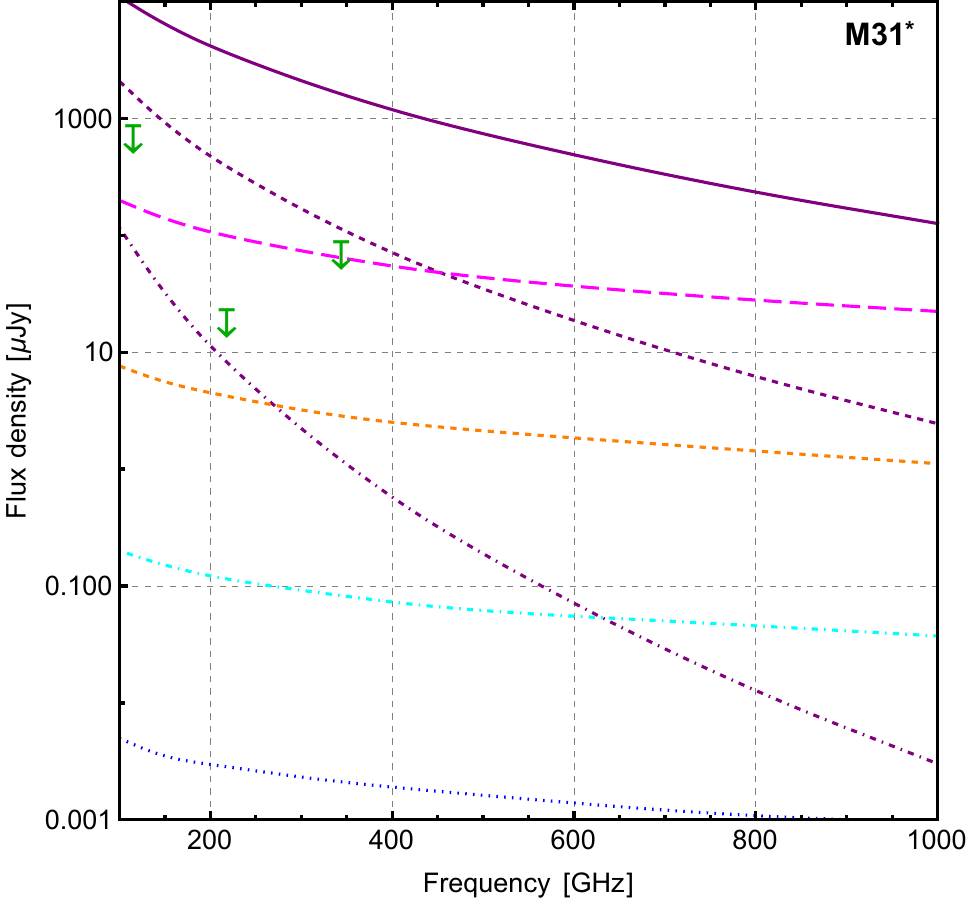}
	\hfill
	\includegraphics[width=0.486\textwidth]{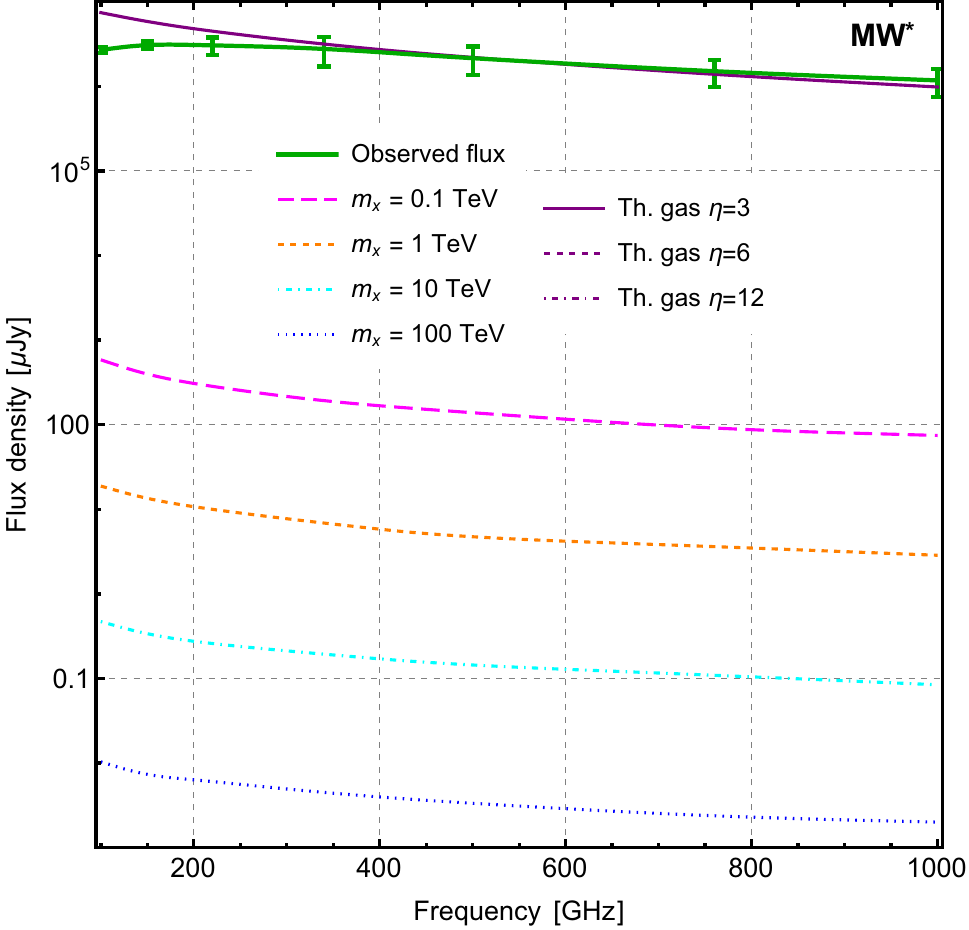}
	\caption{\label{fig:sp}The computed emission spectra from accretion flow and annihilating WIMPs for M31* and MW*. The green data points reflect the observed spectrum for MW* \cite[figure 4]{2024A&ARv..32....3G} and flux density upper limits (3$\sigma_s$) for M31* \cite[table 2]{2025A&A...693A..24M}. DM parameters are the following: annihilation cross section is set to the thermal value $\langle\sigma v\rangle_{th}\approx2.1\cdot10^{-26}$ cm$^3$/s, $\chi\chi\rightarrow0.5b\bar{b}+0.5\tau^+\tau^-$, MED halo density $\rho_0,~\gamma=1.75$.  }
\end{figure}

Now we can test the generic accretion flow model outlined above by comparison of its flux prediction with observational data for MW*. The thermal magnetobremsstrahlung emission spectrum was calculated by integration of the emission coefficient \eqref{eq:jm} (i.e. medium emissivity) over the flow volume. Eqs. \eqref{eq:B}--\eqref{eq:T} enter eq. \eqref{eq:jm} together with all MW* parameter values from table \ref{tab:par}. The result is shown in figure \ref{fig:sp} (right) together with the measured spectrum. The ion-to-electron temperature ratio was set to $\eta=3$ according to the simple recipe in \cite{2022ApJ...930L..16E}. Another tricky detail in this calculation is the choice of inner flow radius. The emission flux due to DM annihilation has very mild sensitivity to this radius. But the thermal emissivity, in contrast, is highly concentrated inside few $R_\bullet$. Hence, the thermal flux is much more sensitive to this choice (to be discussed more below). I lowered the inner boundary of thermal emissivity integration down to $2.5R_\bullet$, which is an effective optimized value matching observations. In principle, photons may arrive from radii down to the photon sphere at $1.5R_\bullet$. However, a little fraction of such photons has a relevance due to huge redshift.

Going back to figure \ref{fig:sp} (right), we see very good agreement between the calculated and observed MW* spectra. They slightly diverge below $\nu\approx200$ GHz due to unaccounted absorption described above. Also, the measured spectrum is uncertain due to its temporal variability. Nevertheless, we can assert a validity of the outlined flow model: it reasonably reproduces both -- MF strength and emission spectrum in MW*.

Another very important application of the flow model is to estimate the thermal emission in M31* and then compare it with the emission due to DM. We know much less about M31*, partially because EHT did not study it yet contrary to MW* and M87*. Microwave data on M31* is rather scarce overall: it was detected at $\nu\leqslant20$ GHz \cite{2017ApJ...845..140Y}, and only shallow flux density upper limits exist above 100 GHz \cite{2025A&A...693A..24M} -- they are shown in figure \ref{fig:sp} (left). Direct model calibration by low frequency data is problematic, since the optically thick regime works there, as was outlined above. Such regime requires a full solution of the emission transport equation. So far I intend to obtain the first basic estimates of sensitivity to WIMPs. And I work in easier regime of optically thin medium. Thus, I calculated essentially the upper limits on thermal emission flux density using $\eta$ as free model parameter. This parameter is the most uncertain and influential at the same time. It can vary over two orders of magnitude \cite{2022ApJ...930L..16E}. Substitution of MW* value $\eta=3$ together with all other M31* model parameter values from table \ref{tab:par} produces the spectrum shown by the continuous purple line in figure \ref{fig:sp} (left). This spectrum greatly exceeds the observed limits. And the absorption can not "defend" such $\eta$ value, because even for the smallest possible $\eta=1$ the optical depth \eqref{eq:tau} is negligible and decreases further with $\eta$ increase at this frequency range. Therefore, it is possible to constrain $\eta$ through the straightforward requirement $F_m(\nu,\eta)=\int j_m(\nu,\eta)d^3R/(4\pi d^2)\leqslant F_\text{obs}(\nu)$. This yields $\eta\leqslant11\div13$ from the most sensitive data point $F_\text{obs}(\nu=220\text{ GHz})\leqslant24~\mu$Jy. Thus, M31* accretion flow has to be significantly colder than that in MW*. The corresponding constraints on the electron temperature in absolute units and optical depth are given in table \ref{tab:par}. An example of corresponding thermal emission spectrum is shown in figure \ref{fig:sp} (left) by the purple dot-dashed line for $\eta=12$. We can also try to do a very rough extrapolation of the emission model down to $\nu=20$ GHz, which is the largest frequency with detected flux: $F_m(\nu=20\text{ GHz},\eta=11)\sim\exp(-\tau_m)\int j_md^3R/(4\pi d^2)\approx50~\mu$Jy, which is quite close to $F_\text{obs}(\nu=20\text{ GHz})=(17\pm3)\mu$Jy \cite[table 2]{2017ApJ...845..140Y}. This indicates the model correctness.

Summarizing this section, it built the basic model of accretion flow medium and tested model validity using observational data for MW* and to some extent for M31*. The following key implications can be drawn. M31* accretion flow has significantly lower MF, electron density and temperature in comparison with MW*. These provide outstanding advantages for WIMP searches. At first, the thermal emission is much fainter in M31*, hence it creates much lower nuisance background for the potential WIMP signal. Secondly, the thermal absorption of synchrotron due to DM is much weaker too, enabling WIMP search over much wider frequency window in optically thin regime. Let us proceed to the study of possible WIMP signal.

\section{\label{sec:tr}Solution of the transport equation}

Having the model of accretion flow medium, we can work out a solution of the transport equation \eqref{eq:f}, which would provide the equilibrium distribution of spectral concentration of $e^\pm$ from DM annihilation. This equation encompasses various physical processes, which define $e^\pm$ diffusion in phase space. A relevance and importance of each process can be judged by its characteristic timescale in comparison with others. Below I elaborate all the processes individually.

\textbf{Diffusive reacceleration} is defined by $D_{pp}$ and represents a fine and model-dependent effect. \cite{2025PhRvL.135l1001C} checked an irrelevance of this effect (eq. (S5) there). I neglect it too with confidence, since even if some reacceleration is present, it can only increase the synchrotron emission flux. Thus, $D_{pp}=0$ provides a conservative case.

\begin{figure}[t]
	\centering
	\includegraphics[width=0.495\textwidth]{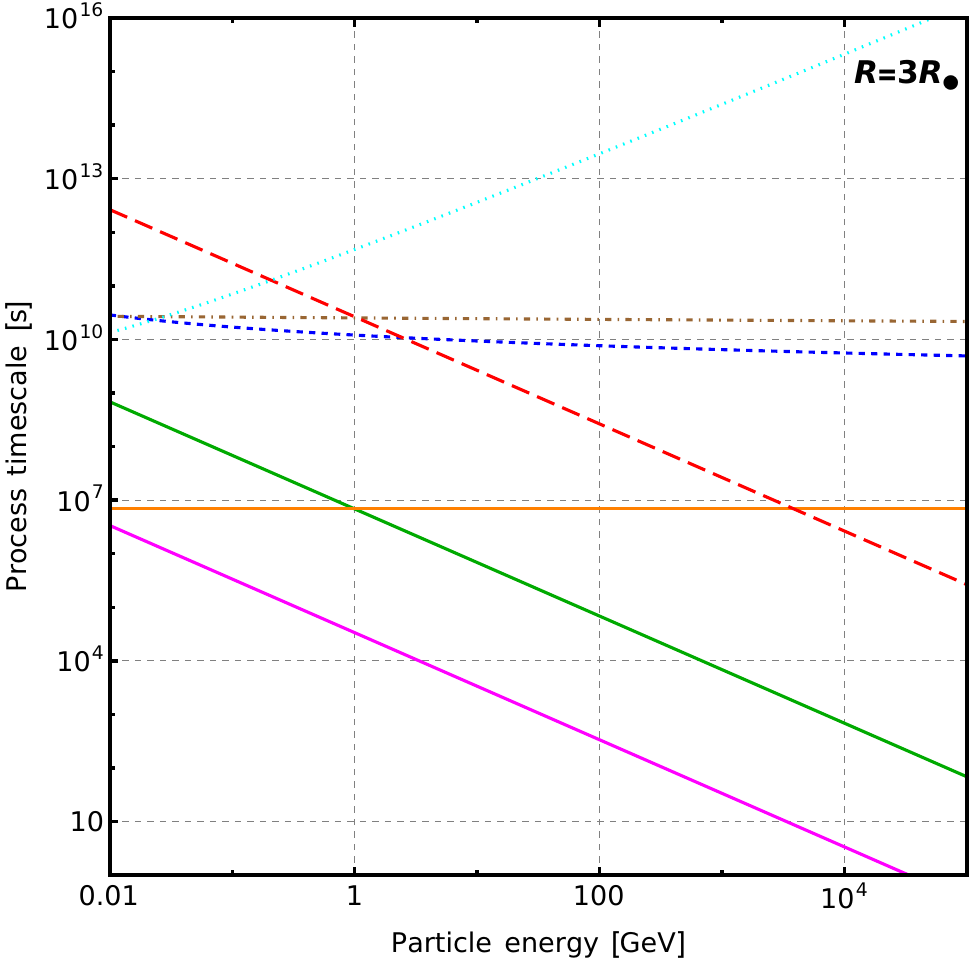}
	\hfill
	\includegraphics[width=0.495\textwidth]{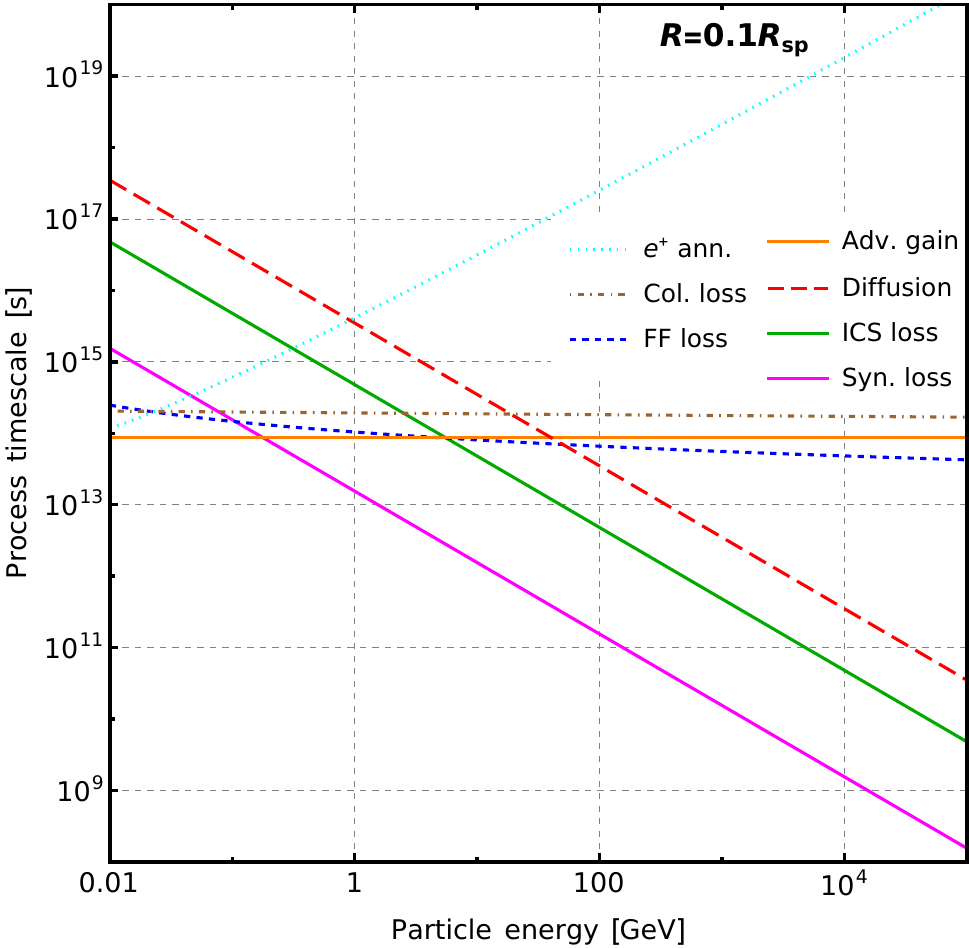}
	\caption{\label{fig:time}Characteristic timescales of various processes, which govern DM $e^\pm$ propagation in M31* accretion flow, as function of energy. The left and right panels reflect the inner and outer edges of the considered emission generation region. The flow model configuration is MED (described in section \ref{sec:flow}). }
\end{figure}

\textbf{Spatial diffusion} is defined by the coefficient $D(R,E)$ and needs to be analyzed more carefully. \cite{2016ApJ...822...88K} investigated an acceleration of non-thermal particles in accretion flow and their diffusion. The latter was obtained to be significantly anisotropic with $D\sim10D_\text{Bohm}$ as characteristic value, where $D_\text{Bohm}=Ec/(3eB(R))$ is Bohm diffusion coefficient. The corresponding process timescale can be roughly estimated as $t_D(R,E)\approx R^2/(6D(R,E))$ \cite[eq. (13)]{1998APh.....9..227C}. It is plotted in figure \ref{fig:time} as function of energy (the red dashed line there) near the inner and outer edges of M31* DM density spike together with timescales for other processes.

\textbf{Adiabatic energy gain} is caused by the flow mechanical compression and can be determined from the generalized first law of thermodynamics, which takes into account gas magnetization: $\dot{p}_a = \dot{E}_a/c=\dot{W}_a/(Nc)=-\zeta\text{\textit\textpeso}\dot{\text{V}}/(Nc)$, where $W,~N$, \textit\textpeso, V are the internal energy, particle count, pressure and volume of $e^\pm$ gas respectively. $\zeta=1\div1.25$ reflects the effect of magnetization -- its thorough derivation can be seen in \cite[section 3.1]{2005A&A...440..223B}. All the previous studies, which were mentioned in section \ref{ssec:i}, ignored this effect setting $\zeta=1$, which corresponds to the case \textit\textpeso$_\text{MF}\ll$ \textit\textpeso. However, in reality, it is likely vice versa: \textit\textpeso$_\text{MF}\gg$ \textit\textpeso~ $\rightarrow \zeta=1.25$. Nevertheless, this correction has a negligible effect on the synchrotron emission flux: the latter increases by $\lesssim10\%$, if $\zeta$ increases from 1 to 1.25 (MED configuration). And I set $\zeta=1$ for simplification of transport equation solving. Then taking into account the equation of state for ultrarelativistic gas \textit\textpeso = $W$/(3V) and $\dot{\text{V}}/$V = $\nabla\vec{V}$, it is easy to get
\begin{equation}\label{eq:pa}
\dot{p}_a = -\frac{p}{3}\nabla\vec{V} = -\frac{p}{3}\frac{1}{R^2}\frac{\partial}{\partial R}(R^2V_R) = -\frac{pV_R}{2R},
\end{equation}
where I assumed $V_\theta=0,~V_\varphi=\text{const}(\varphi)$ and used eq. \eqref{eq:V}. The corresponding heating timescale can be estimated as $t_a\sim p/\dot{p}_a=E/\dot{E}_a$, it does not depend on momentum and is shown by the orange line in figure \ref{fig:time}.

\textbf{Energy losses} $\dot{p}_r$ are mainly radiative and are defined, in general, by four processes: synchrotron, inverse Compton scattering (ICS) and bremsstrahlung emissions; and Coulomb collisional losses. The expressions for loss rates can be found in e.g. \cite{Ginzburg}. All four loss rate timescales $p/\dot{p}_r$ were calculated using eqs. \eqref{eq:B}, \eqref{eq:ne}, \eqref{eq:U} and are plotted in figure \ref{fig:time}. We see that the bremsstrahlung and Coulomb losses are much slower and, hence, negligible in comparison with the synchrotron and ICS losses almost everywhere in the phase space. Therefore I included only the latter two processes in the model through the following expression:
\begin{equation}\label{eq:pr}
\dot{p}_r = -\frac{32\pi e^4}{9m_e^2c^4}\left(\frac{B^2(R)}{8\pi}+U(R)\right)\left(\frac{p}{m_ec}\right)^2.
\end{equation}
Meanwhile, the authors \cite{2004JCAP...05..007A,2025PhRvL.135l1001C} missed the main component of radiation field $U(R)$, which is described by the first term in eq. \eqref{eq:U}: $L/(3\pi cR^2)\gg U_\star$. This did not affect their results just because even the full correct ICS loss rate (the green line in figure \ref{fig:time}) is subdominant with respect to the synchrotron loss rate (the magenta line in figure \ref{fig:time}).

\textbf{Positron annihilation} may degrade the positron concentration. Hence we need to estimate the positron lifetime and compare it with other timescales. The lifetime can be estimated as the time until collision with a thermal electron: $t_+\approx1/(n_e(R)\sigma_+(E)c)$, where $\sigma_+$ is the annihilation cross section. The latter is dominated by the direct in-flight process at the energies of interest. The corresponding cross section can be estimated as $\sigma_+(E)\approx\pi e^4/(m_e^2c^4(1+E/(m_ec^2)))(\ln(2(1+E/(m_ec^2)))-1)$ \cite[eqs. (88.8)-(88.9)]{LL4}. Substitution of the electron concentration \eqref{eq:ne} yields the lifetime shown by the cyan dotted line in figure \ref{fig:time}. This annihilation rate is much slower than the radiative energy loss rate. Thus, the positrons from DM annihilation survive completely while they generate the synchrotron emission of interest.

Now we can compare the rates of all processes. The radiative energy losses \eqref{eq:pr} is clearly the leading process in the innermost region of DM spike. In outer regions the role of adiabatic gain becomes somewhat important too. Considering the whole spike, we can neglect by all other processes. This assumption may be ambiguous only at the lowest energies $\sim(10\div100)$ MeV. However, this first energy decade has a very little contribution to the emission flux. Thus, overall, even the simplest possible solution of the transport equation \eqref{eq:f}, which takes into account the radiative losses only, should provide an adequate approximation of concentration in phase space. Such solution is well-known and has the form (in energy representation, implying spherical symmetry):
\begin{equation}\label{eq:psi0}
\psi(R,E) = -\frac{1}{\dot{p}_rc}\int\limits_E^{m_xc^2}q(R,E')dE'.
\end{equation}
Nevertheless, I included in the model the second-order effects due to the medium motion and compression for a sake of accuracy. In this case solving the transport equation becomes substantially more difficult. Its brief elaboration follows. Setting $D=D_{pp}=0$ and substituting \eqref{eq:pa} in \eqref{eq:f} yields
\begin{equation}\label{eq:fr}
\vec{V}\nabla f+(\dot{p}_r+\dot{p}_a)\frac{\partial f}{\partial p}+\frac{4\dot{p}_r}{p}f = q_p.
\end{equation}
This partial differential equation can be transformed into ordinary one and further solved with help of the ancillary equation, which describes the particle radial motion:
\begin{equation}\label{eq:dp/dR}
\frac{dp}{dR}\equiv\frac{dp}{dt}\frac{dt}{dR} = \frac{\dot{p}_r+\dot{p}_a}{V_R}.
\end{equation}
Then substitution of the straightforward relation $\frac{df}{dR}=\frac{\partial f}{\partial R}+\frac{dp}{dR}\frac{\partial f}{\partial p} = \frac{dp}{dR}+\frac{\dot{p}_r+\dot{p}_a}{V_R}\frac{\partial f}{\partial p}$ into eq. \eqref{eq:fr} yields
\begin{equation}\label{eq:ff}
\frac{df}{dR}+\frac{4\dot{p}_r}{pV_R}f = \frac{q_p}{V_R}.
\end{equation}
This linear nonhomogeneous equation of the first order can be solved easily, treating $R$ as independent variable:
\begin{equation}\label{eq:fs}
f(R,p) = \int\limits_{R_{sp}}^R\exp\left(-\int\limits_{R_0}^R\frac{4\dot{p}_r'}{p'V_R(R')}dR'\right)\frac{q_p(R_0,p_0)}{V_R(R_0)}dR_0,
\end{equation}
where the boundary condition $f(R_{sp},p)=0$ was implemented. We can solve the integral under exponent using the relation $\dot{p}_r'=dp'/dt-\dot{p}_a'$ and eq. \eqref{eq:pa}: $\int\limits_{R_0}^R\frac{4\dot{p}_r'}{p'V_R(R')}dR' = \int\limits_{R_0}^R\frac{4}{p'V_R(R')}\left(\frac{dp'}{dt}+\frac{p'V_R(R')}{2R'}\right)dR' = 4\ln\frac{p}{p_0}+2\ln\frac{R}{R_0}$. Substitution of this expression into eq. \eqref{eq:fs} provides:
\begin{equation}\label{eq:fsf}
f(R,p) = \int\limits_{R_{sp}}^R\left(\frac{p_0}{p}\right)^4\left(\frac{R_0}{R}\right)^2\frac{q_p(R_0,p_0)}{V_R(R_0)}dR_0.
\end{equation}
The physical meaning of this solution is that the concentration $f(R,p)$ is formed by all the particles with initial momenta $p_0$ created at all radii $R_0$ above $R$. $p_0$ needs to be expressed through the independent variables, i.e. $p_0=p_0(p,R,R_0)$. This is done by solving \eqref{eq:dp/dR} for $p(p_0,R,R_0)$ and then extracting $p_0$ from this solution. Using eqs. \eqref{eq:B}, \eqref{eq:V}, \eqref{eq:U}, \eqref{eq:pa} and \eqref{eq:pr}; eq. \eqref{eq:dp/dR} can be expressed as 
\begin{equation}\label{eq:dp/dR-B}
\frac{dp}{dR}+\frac{p}{2R} = (AR^{-3/2}+A_\star R^{1/2})p^2,
\end{equation}
where I approximated the complex piecewise MF profile \eqref{eq:B} by the simplified profile $B(R) = B(3R_\bullet)3R_\bullet/R$; which follows the original profile closely, reflects the basic fact of magnetic flux conservation and eases solving of \eqref{eq:dp/dR}. Eq. \eqref{eq:dp/dR-B} belongs to the well-known class of Bernoulli differential equations. Its solution and subsequent $p_0$ extraction give:
\begin{equation}\label{eq:p0}
	\begin{split}
p_0 &= \frac{pR\sqrt{R_0}}{R_0\sqrt{R}+p(R-R_0)(A+A_\star RR_0)}, \\ A &= -\frac{32\pi e^4}{9m_e^4c^6V_R(1\text{cm})}\left(\frac{(B(3R_\bullet)3R_\bullet)^2}{8\pi}+\frac{L}{3\pi c}\right),~A_\star = -\frac{32\pi e^4U_\star}{9m_e^4c^6V_R(1\text{cm})}.
	\end{split}
\end{equation}
Finally, we need to transform the concentration in phase space into the energy representation. This is done by the following relations, which reflect momenta isotropy: $q_p(R,p)=q(R,E=pc)c/(4\pi p^2),~\psi(R,E)=f(R,p=E/c)4\pi p^2/c$. Putting these relations and (partially) \eqref{eq:p0} into eq. \eqref{eq:fsf} provides the final solution:
\begin{equation}\label{eq:psi}
\psi(R,E) = \int\limits_{R_{sp}}^R\frac{q(R_0,p_0(E/c,R,R_0)c)R_0^3}{(R_0\sqrt{R}+E(R-R_0)(A+A_\star RR_0)/c)^2V_R(R_0)}dR_0.
\end{equation}
For information, appendix \ref{sec:a} shows the generalized solution with $\zeta\neq1$.

\section{\label{sec:obt}Obtained emission due to WIMP annihilation}

Combining together eqs. \eqref{eq:F}, \eqref{eq:f}, \eqref{eq:p0} and \eqref{eq:psi}; we can write out the key equation for the synchrotron emission flux density due to annihilating DM:
\begin{equation}\label{eq:Ff}
F(\nu,R_m) = \frac{\langle\sigma v\rangle}{(dm_x)^2}\int\limits_{3R_\bullet}^{R_m}\int\limits_{R_{sp}}^R\int\limits_{E_0}^{m_xc^2}\frac{P_e(\nu,R,E)\rho^2(R_0)R^2R_0^3dRdR_0dE}{V_R(R_0)(R_0\sqrt{R}+E(R-R_0)(A+A_\star RR_0)/c)^2}\frac{dN_e}{dE}(p_0c),
\end{equation}
where $p_0=p_0(E/c,R,R_0)$ comes from eq. \eqref{eq:p0}; $dN_e/dE=0$, if $E_0>p_0c>m_xc^2$; $R_m\leqslant R_{sp}$.

\begin{figure}[t]
	\centering
	\includegraphics[width=0.495\textwidth]{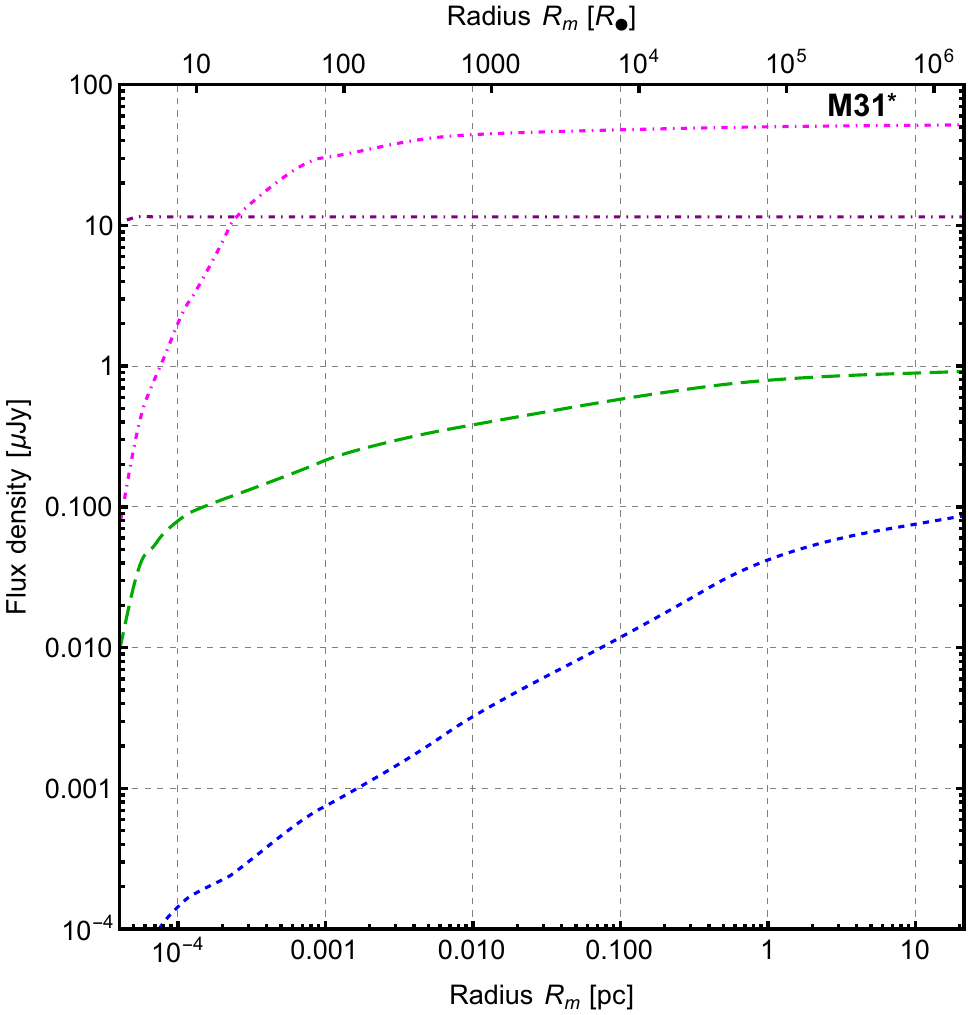}
	\hfill
	\includegraphics[width=0.495\textwidth]{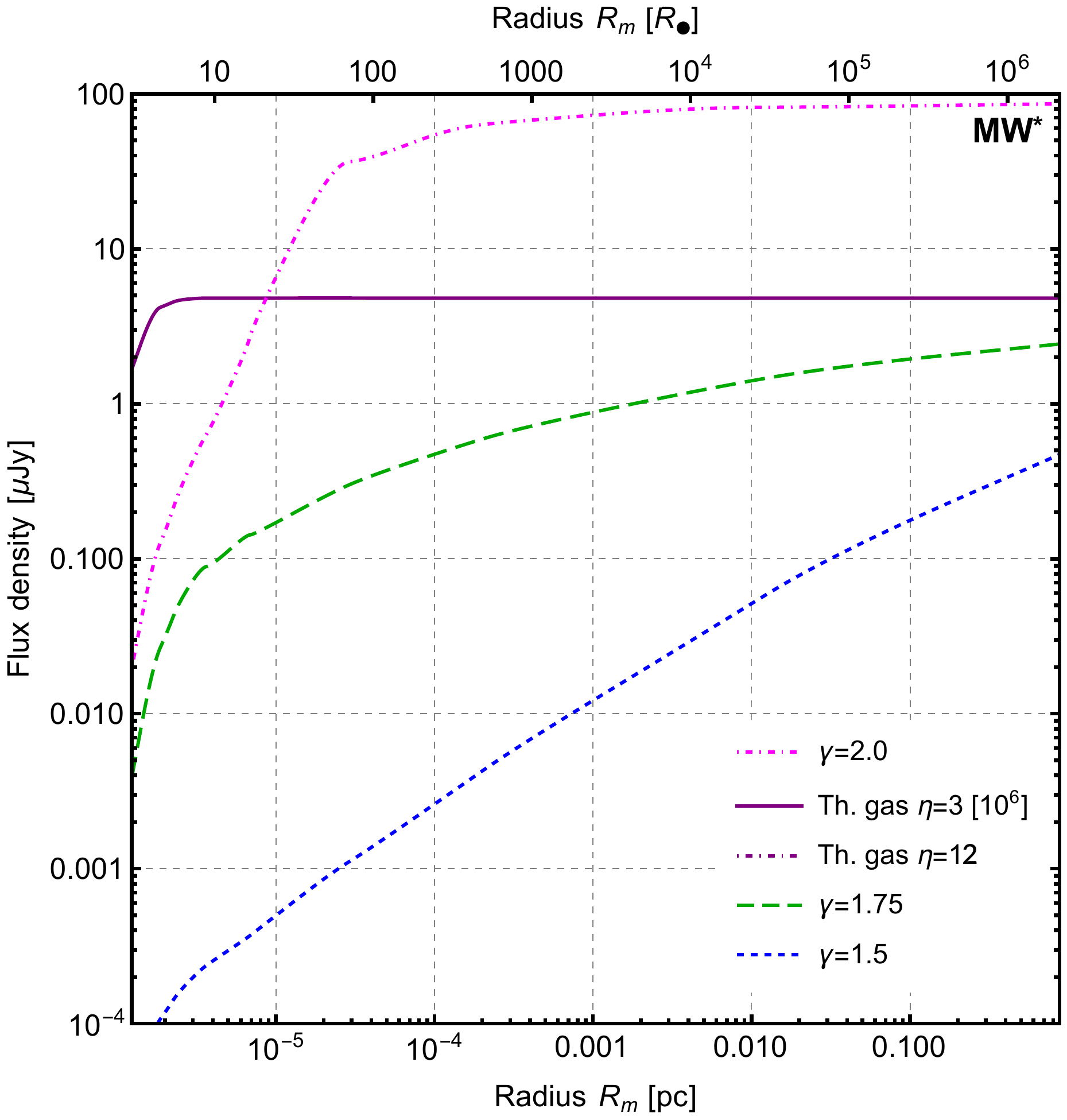}
	\caption{\label{fig:F(R)}The flux density dependence on the outer radius of DM spike emission integration region $F(\nu=200\text{ GHz},R_m)$ for M31* (left panel) and MW* (right panel) for various values of spike density slope $\gamma$. The thermal emission profile is also shown by the purple lines (note that MW* profile is reduced by $10^6$ times). The annihilation cross section is thermal, $m_x$ = 3 TeV, $\chi\chi\rightarrow0.5b\bar{b}+0.5\tau^+\tau^-$, MED halo density $\rho_0$. The range of horizontal axis is $3R_\bullet\div R_{sp}$.}
\end{figure}

Let us start the analysis of emission due to DM from studying the contributions of various spike radial layers into the total flux. Figure \ref{fig:F(R)} serves for this purpose: it shows the flux density \eqref{eq:Ff} as function of the radius of emission integration region $R_m$ inside the spike at frequency 200 GHz, where the absorption becomes small in both objects. This figure also contains the same dependence for the thermal (magnetobremsstrahlung) emission flux density $F_m$ produced by the accretion flow. First of all, we see a huge difference in the latter flux between MW* and M31* (MW* thermal flux is plotted there in [Jy]), but the fluxes due to DM differ in contrary very slightly. Thus, essentially, M31* DM signal to thermal background ratio is expected to be larger by $\sim10^5$ than that for MW*! Moreover, as was explained above, M31* thermal flux estimate represents just an upper limit rather than solid measurement, as it is for MW*. Thus, we have unveiled a truly fascinating value of M31* for WIMP searches. One may naturally doubt, why the difference in signal-to-background ratio is so large. I see at least two key reasons for that. One is large difference in SMBH masses: the spike emission flux grows with the mass increase very steeply. Another reason is substantially lower anticipated MF strength in M31*: the former enables much slower cooling rate of DM $e^\pm$; hence, they live and "shine" much longer. 

Figure \ref{fig:F(R)} demonstrates also an important common difference in emissivity distribution between the flow and DM spike: all the flow emission is generated inside very thin layer up to few $R_\bullet$, while the spike emission gets accumulated from much larger spatial scale. This difference is useful for observational separation of two emission components. In principle, it could be possible to mask a bright thermal emission, which is localized in a very small vicinity around the center, in order to conduct a clear imaging of the potential DM spike emission. However, in practice it could be difficult, since it would require very long mosaic imaging with a very small telescope beam. In general, it is important to keep in mind, that figure \ref{fig:F(R)} can not demonstrate precisely the dependence of flux on the angular radius of integration region on the sky. Calculation of such dependence requires integration of the spike emissivity over two cylindrical coordinates $\lbrace r,z \rbrace$ (for illustration one can see e.g. \cite[section IID]{2013PhRvD..88b3504E}) rather than over one spherical coordinate $R$ in the presented case. This would add another dimension into the multiple integral \eqref{eq:Ff}, which complicates computations substantially. Thus, I left the detailed analysis of surface brightness (i.e. spike emission intensity) distribution on the sky for future work. Here I made approximate estimates of the emission morphology using spherical integration.

The flux density depends expectedly strongly on the spike density profile slope $\gamma$. And another important implication from figure \ref{fig:F(R)} is that the outermost spike layer $0.1R_{sp}\lesssim R \lesssim R_{sp}$ contributes very little to the total flux. At the same time, telescopic imaging of the whole spike, i.e. when $R\leqslant R_{sp}$, may grasp much more nuisance emission and noise in comparison with much smaller imaging region with $R\leqslant 0.1R_{sp}$. For this reason, I chose to set $R_m = 0.1R_{sp}$ for the whole analysis here, since larger $R_m$ would likely degrade signal-to-noise ratio.

Now let us look at the spectral properties of DM spike emission. They are illustrated at figure \ref{fig:sp} for four characteristic WIMP masses. We see a strong decrease of the emission flux density with mass increase, which is a very common behavior. Another important property is quite slow flux decrease with frequency, i.e. a hard spectrum, especially at higher frequencies. The thermal flow emission spectrum is much softer in M31*. This is caused by the low thermal electron temperature $\eta\lesssim10$: the former have characteristic energy about MeV in contrast to DM $e^\pm$, which have GeV-TeV energies. Such difference provides another important utility for distinguishing two emission components. Thus, multifrequency emission detection may identify its source quite unambiguously by the spectral slope. A remarkable similarity between MW* and M31* emission fluxes due to DM is caused likely just by a lucky combination of parameters of those SMBHs. But a large prevalence of the thermal emission in MW* over that due to DM leaves poor chances for WIMP detection in MW*, at least for realistic $\gamma$ values.

\begin{figure}[t]
	\centering
	\includegraphics[width=0.5\textwidth]{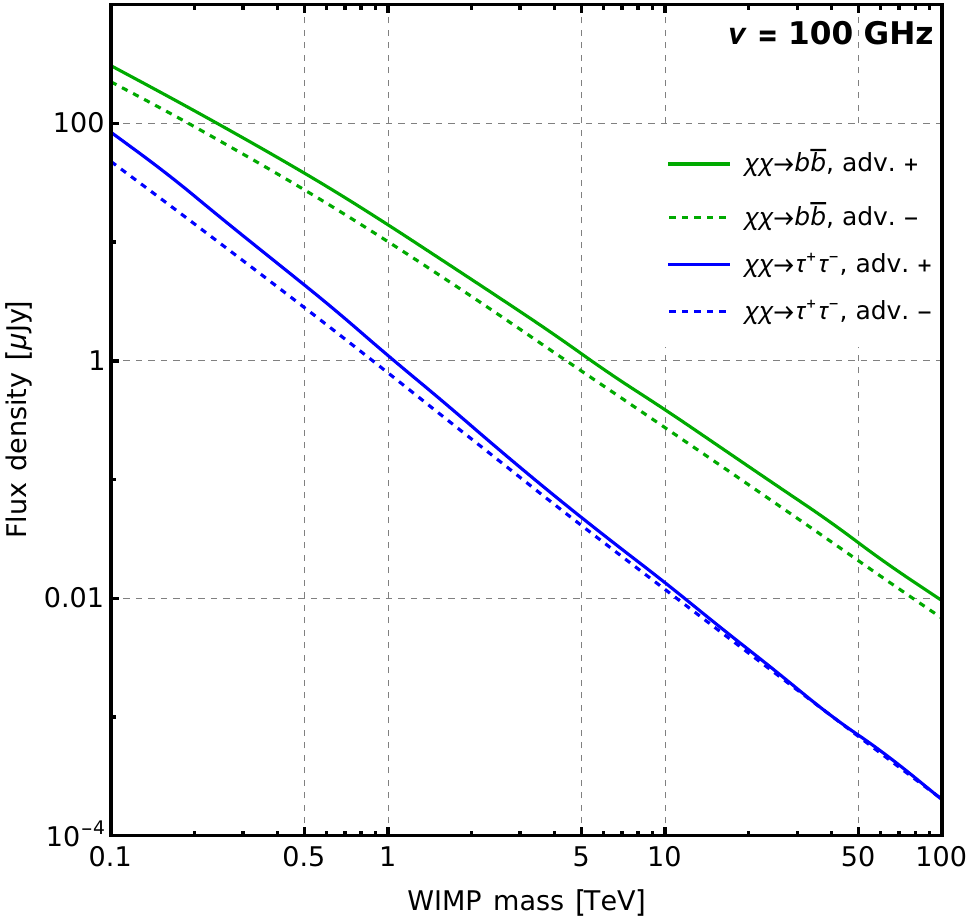}
	\caption{\label{fig:adv}DM spike emission flux density dependence on WIMP mass at frequency 100 GHz for both considered annihilation channels with (the solid lines, $e^\pm$ distribution from eq. \eqref{eq:psi}) and without (the dashed lines, $e^\pm$ distribution from eq. \eqref{eq:psi0}) effects of accretion flow radial motion. The annihilation cross section is thermal, MED halo density $\rho_0$, $\gamma=1.75$.}
\end{figure}

It is very useful to compare the emission fluxes calculated from different versions of $e^\pm$ spectral concentration, which are expressed by solutions \eqref{eq:psi0} and \eqref{eq:psi}. Such comparison provides model self-check and also assesses the magnitude of effects due to flow motion (i.e. when $V_R\neq0$). These effects influence the emission flux in two opposite ways: the adiabatic compression reheats cooling $e^\pm$ and, hence, increases the flux; the radial motion may infall $e^\pm$ into SMBH before they radiate out their energy and, hence, decreases the flux. Figure \ref{fig:adv} illustrates the difference between static ($V_R=0$) and dynamic cases ($V_R\neq0$), as well as full flux dependence on WIMP mass. We see that the adiabatic compression effect dominates for both channels. However, the overall flux increase due to both effects is rather modest, although still significant. A relative proximity of emission fluxes in two cases validates a basic correctness of more complex solution \eqref{eq:psi}.

\section{\label{sec:res}ALMA sensitivity to WIMP signal}

Considering all existing interferometric telescopes for millimeter wavelengths, ALMA is probably the most sensitive and suitable for our purposes. This section provides ALMA sensitivity estimates to WIMP annihilation cross section depending on WIMP mass and spike density profile. At the beginning, we need to choose optimal frequency and imaging mode/resolution for prospective deep observations of M31*. So far ALMA observed this object during just $\epsilon\approx1$ hour (h) at $\nu\approx350$ GHz \cite{2025A&A...693A..24M}, which is not a good frequency for DM search. I used the official ALMA sensitivity calculator\footnote{\url{https://almascience.nrao.edu/proposing/sensitivity-calculator}} for selection of imaging parameters. Basic instrumental parameters were set to the following common values: dual polarization, 7.5 GHz bandwidth per polarization, default choice of atmospheric water vapor profile, 50 (out of 50) 12m antennas are on.

As was mentioned above, no emission from M31* has been seen so far. Considering this fact, I view it sensible to observe the whole spike or/and accretion flow at 2-3 frequencies without attempts to resolve its internal structure at the next stage of ALMA observations with larger exposure. Potential detection will allow to assess (at least, partially) the emission origin just by the spectral shape, as was outlined in the previous section. Such preliminary assessment will give hints for further strategy: whether chances to discover DM signal exist and, therefore, much larger mosaic exposure is needed for high-resolution spike intensity mapping. Thus, I estimated ALMA sensitivity to WIMPs assuming the simple observations with single synthetic beam, which matches the spike angular radius $R_m/d=0.1R_{sp}/d\approx0.35''\div1.3''$ (from table \ref{tab:par}). ALMA easily provides such beam sizes at relevant frequencies. 

\begin{figure}[t]
	\centering
	\includegraphics[width=0.495\textwidth]{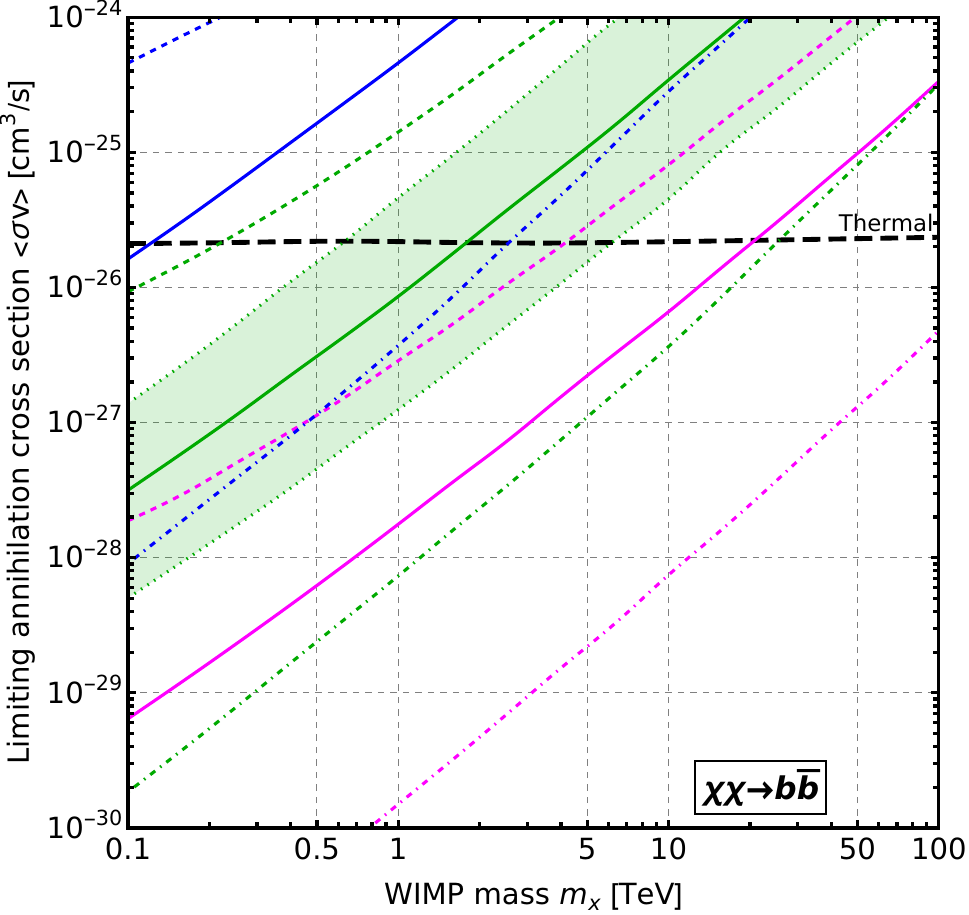}
	\hfill
	\includegraphics[width=0.495\textwidth]{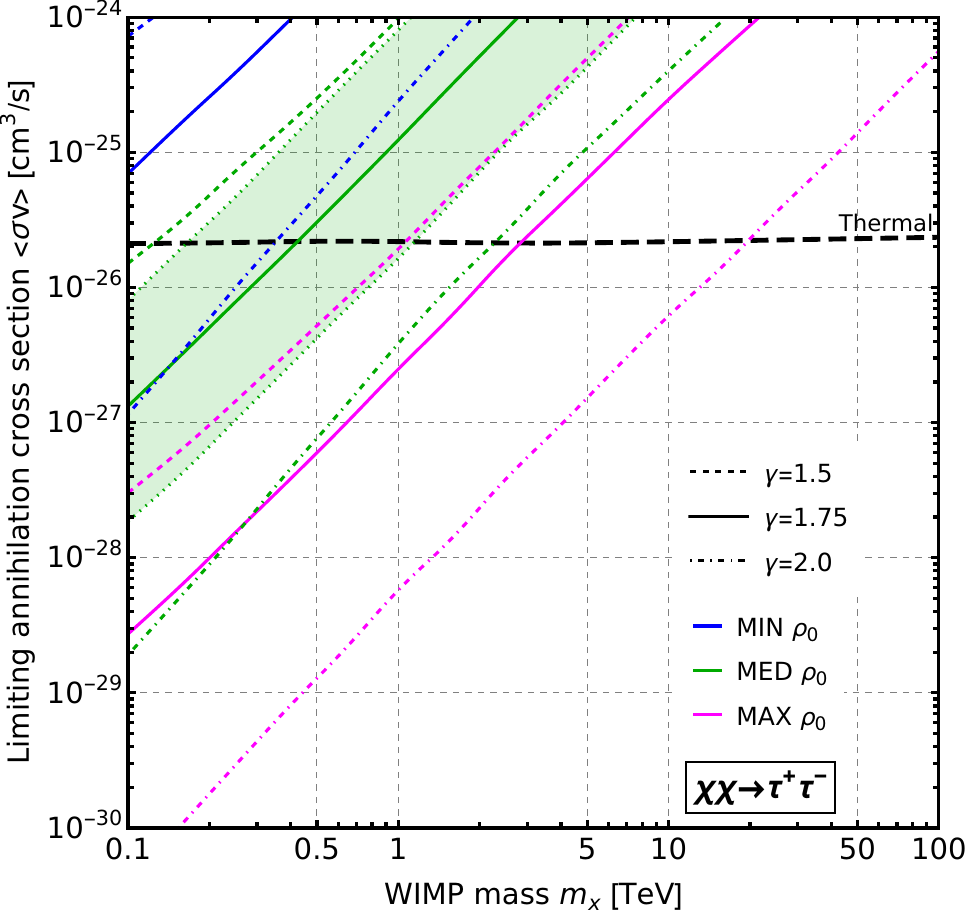}
	\caption{\label{fig:sv}The limiting WIMP annihilation cross section vs mass, which can be probed by ALMA observations of M31* at $\nu=100$ GHz with the exposure time $\epsilon=24$ h, for both annihilation channels and denoted spike density profile parameter configuration. The sensitivity lines are obtained with MED parameters of SMBH and accretion flow medium from table \ref{tab:par}. The shaded green belts show the systematic model uncertainty for MED $\lbrace \gamma,\rho_0 \rbrace$ configuration, if $\lbrace M,\sigma_c,\alpha,\beta,\delta \rbrace$ are varied over their MIN-MAX ranges. The horizontal black dashed line denotes the thermal relic annihilation cross section $\langle\sigma v\rangle_{th}(m_x)$ from \cite{2020JCAP...08..011S,sv}.}
\end{figure}

The choice of optimal frequency is based on both emission and sensitivity spectra. The latter, as the instrumental noise level inside the beam, does not depend on beam size (defined by configuration of antenna baselines) and reaches minimum (i.e. the best sensitivity) at $\nu\approx100$ GHz. Thus, it does not make sense to go above this frequency, since both DM signal intensity and telescope sensitivity decreases there, unless we need to avoid too bright thermal emission. Below 100 GHz the picture is not so clear: absorption may develop, the thermal background arises steeply, and decrease of sensitivity may overtake an increase of signal. I chose naturally 100 GHz for calculation of sensitivity to WIMP parameters. And I considered the idealized case of no confusion from the thermal emission at this frequency in order to evaluate the full ALMA potential. I obtained the sensitivity limits for the exposure time $\epsilon=24$ h, which provides the instrumental noise level $3\sigma_n(\nu=100\text{ GHz})\approx6~\mu$Jy according to the sensitivity calculator. The limiting annihilation cross section, which can be probed by such observation, was computed from the simple relation:
\begin{equation}\label{eq:sv}
F(\nu=100\text{ GHz},\langle\sigma v\rangle_\text{lim},m_x,\gamma,\rho_0,...)=3\sigma_n(\nu=100\text{ GHz}),
\end{equation}
which implies 99.7\% statistical confidence level. This calculation is rather non-trivial, since the flux density $F$ expressed by eq. \eqref{eq:Ff} depends on $\langle\sigma v\rangle$ non-linearly due to dependence of DM density profile $\rho(R)$ on $\langle\sigma v\rangle$. The latter dependence arises in the case of formation of the inner flattened spike core due to DM annihilation -- can be seen in \cite[eqs. (2.2)-(2.3)]{2026JCAP...04..002E}. Thus, for each model parameter configuration, I computed $F(\langle\sigma v\rangle,m_x)$ at the discrete $25\times16$ grid of cross section and mass values, which were restricted by the ranges ($10^{-30}\div10^{-24}$) cm$^3$/s and $(0.1\div100)$ TeV respectively. Then I interpolated the obtained 3D array and derived machinely the solution $\langle\sigma v\rangle_\text{lim}(m_x)$ of eq. \eqref{eq:sv}. These solutions are presented in figure \ref{fig:sv} for both considered annihilation channels, 3 MIN-MAX $\gamma$ and $\rho_0$ values.

We can outline the following main properties of obtained results. Variation of $\gamma$ and $\rho_0$ over the allowed MIN-MAX ranges translates into variation of $\langle\sigma v\rangle_\text{lim}$ over more than six orders of magnitude. Such huge range is generally expected, because DM annihilation rate inside the spike strongly depends on the density profile (eq. \eqref{eq:f}). Thus, modest spike densities can not produce detectable emission even for the lightest considered WIMPs. In contrast, very dense spike profiles are able to unveil even the heaviest possible WIMPs around the unitarity mass limit, at least for certain annihilation channels and their combinations. Considering MED, i.e. the most realistic, $\lbrace \gamma,\rho_0 \rbrace$ configuration, we can deduce the following largest reachable thermal WIMP masses for that (the intersection points of green solid and black dashed curves in figure \ref{fig:sv}): $m_x = 1.8$ TeV for $b\bar{b}$ and $m_x = 0.42$ TeV for $\tau^+\tau^-$. Leptonic channels are less constraining than hadronic ones, which is common for the synchrotron emission due to WIMPs. The approximate dependence of limiting cross section on mass is $\langle\sigma v\rangle_\text{lim} \propto m_x^\omega$, where $\omega=1.6$ for $b\bar{b}$ and $\omega=2.0$ for $\tau^+\tau^-$. The power-law index $\omega$ decreases slightly with an increase of DM density.

So far we have discussed the primary systematic model uncertainties due to those in spike density profile. However, the "secondary" uncertainties related to those in SMBH mass $M$, galactic central velocity dispersion $\sigma_c$, accretion flow medium parameters $\alpha,\beta,\delta$; are important too, since they also influence the spike density, as well as MF and flow velocity profiles. I assessed the extent of these uncertainties by variation of $\lbrace M,\sigma_c,\alpha,\beta,\delta \rbrace$ over their MIN-MAX ranges (described in section \ref{sec:flow}) for MED $\lbrace \gamma,\rho_0 \rbrace$ configuration. The resulting variation in $\langle\sigma v\rangle_\text{lim}(m_x)$ is shown by the green shaded zones in figure \ref{fig:sv}. We can note highly symmetric uncertainty distribution around the MED line and modest ($\approx1.5$ orders of magnitude) uncertainty vertical width, which agree with general expectations.

It is important to keep in mind, that figure \ref{fig:sv} represents the particular choice of observation frequency and duration. Longer exposure and further sensitivity increase are realistically possible. The reachable cross section scales with exposure time as $\langle\sigma v\rangle_\text{lim} \propto \epsilon^{-1/2}$, if to approximate $F\propto \langle\sigma v\rangle$, i.e. $\rho(\langle\sigma v\rangle)=$ const. The situation with frequency tuning is less straightforward. Frequency increase degrades the sensitivity quite fast due to double effect of signal fainting and noise increase. Frequency decrease requires the detailed modeling of thermal emission and absorption, which is left for future work. But the presented particular observational scenario provides the first/initial evaluation of an overall perspective of this methodology. This evaluation shows a competitive sensitivity, which is discussed in more details in the next subsection.

\subsection{\label{ssec:comp}Comparison of sensitivity with other methodologies}
Let us judge now how the developed methodology fits into a big picture of the whole field of WIMP indirect searches. Proceeding from the particular to general, let us compare the obtained sensitivity with others related to DM density spikes. First of all, my previous work \cite{2026JCAP...04..002E} estimated CTA sensitivity to the prompt gamma-ray emission in M31* -- shown in figure 5 (left) there. We see generally much better sensitivity to the cross section in the microwave band: roughly by 3.5 orders of magnitude considering, for example, $m_x=3$ TeV, $b\bar{b},~\gamma=1.75$, MAX $\rho_0$ (3 TeV here does not concern the thermal Wino, it is just the representative (geometric) mean over the considered mass range $\sqrt{0.1\cdot100}\approx3$). We may note the following advantages of the gamma-ray band: much smaller slope $\omega$ and milder dependence on annihilation channel. However, these advantages are unlikely able to win, i.e. ALMA observations of M31* look to be much more promising with respect to those by CTA.

Considering other nearby SMBHs, MW* in gamma-ray band was evaluated in \cite{2026JCAP...04..002E} too (see figure 5 (right) there), as well as in other works, e.g. by \cite{2023JCAP...08..063B} recently. The main difficulty there is quite bright astrophysical background, which allows so far very weak WIMP constraints, and it is hard to predict whether CTA observations will be able to disentangle various background components. The microwave synchrotron emission in MW* was estimated in section \ref{sec:obt}: figure \ref{fig:sp} (right) demonstrates, that WIMP signal is tiny relative to the thermal background. An advantage of MW* is its proximity, which allows high-resolution imaging of the spike region. Thus, outer spike layers, i.e. outside the bright thermal halo around SMBH, may potentially be used for WIMP search, although such method requires a dedicated evaluation. As was discussed in section \ref{sec:i}, M87* also attracted significant attention. But \cite{2026JCAP...04..002E} demonstrated, that M87* is expected to have the faintest WIMP gamma-ray signal among nearby SMBHs for realistic spike densities. Regarding the synchrotron emission; we can estimate it using the model developed here, the simplified MF distribution $B(R)=30R_\bullet/R$ G from \cite{2021ApJ...910L..13E}, SMBH parameters and $\rho_0$ from \cite[table 1]{2026JCAP...04..002E}. These all yield the flux density $F(\nu=230\text{ GHz})\lesssim10^{-11}$ Jy for $m_x=3$ TeV, $b\bar{b},~\gamma=1.75$; which is essentially zero in comparison with the observed $F_m(\nu=230\text{ GHz})\sim1$ Jy \cite{2021ApJ...910L..13E}. Thus, M87* can be discarded from consideration in our context quite confidently and completely. The third SMBH after M31* and MW* by WIMP detection potential is likely NGC 3115*, if to judge by \cite[table 1]{2026JCAP...04..002E}.

Besides SMBH observations, probably only one alternative methodology is known for probing heavy TeV-scale WIMPs in a foreseeable future -- namely, CTA observations of the Galactic center region \cite{2021JCAP...01..057A}. It is not easy to compare their sensitivity directly with that obtained here, because \cite{2021JCAP...01..057A} chose the intermediate annihilation channel $\chi\chi\rightarrow W^+W^-$ as the main one in their analysis. For an approximate comparison, I calculated $\langle\sigma v\rangle_\text{lim}(m_x=3\text{ TeV},\chi\chi\rightarrow W^+W^-,\gamma=1.75)$ for MIN-MAX range of $\rho_0$ values in my model, which yielded the range $\langle\sigma v\rangle_\text{lim}\sim(10^{-27}\div10^{-24})$ cm$^3$/s. \cite{2021JCAP...01..057A} obtained CTA sensitivity at the level $\langle\sigma v\rangle_\text{lim}^{MW}\sim(10^{-26}\div10^{-25})$ cm$^3$/s for the same $m_x$ and with similar $\rho_0^{MW}$ variation. The sensitivity uncertainty range for M31* in microwave band is much wider. However, M31* may provide better sensitivity at least \emph{potentially}, i.e. in case of lucky parameter scenario. Thus, both methodologies are competitive and complementary.

\textbf{Higgsino} represents the specific and, probably, most anticipated WIMP, especially in view of very recent hint for its detection by LUX-ZEPLIN experiment \cite{2026arXiv260902823A}, as was pointed out in e.g. \cite{2026arXiv260901583F,2026arXiv260902608D}. Let us estimate Higgsino detection potential by ALMA observations of M31*. The typical thermal Higgsino is expected (from theory) to have the following properties according to \cite{2023PhRvL.130t1001D}: $m_x$ = 1.08 TeV, $\langle\sigma v\rangle = 1.3\cdot 10^{-26}$ cm$^3$/s, $\chi\chi\rightarrow 0.6W^+W^- + 0.4Z^0Z^0$. In the absence of thermal astrophysical background, $3\sigma_n$ detection would require $\epsilon\approx 41$ h at 100 GHz and 47 days at 300 GHz of continuous signal collection (MED configuration). The sensitivity drops very fast with frequency, but lower frequencies are able to provide quite fast and easy detection in good case scenario; especially in comparison with CTA, which needs to collect data during up to $\sim10$ years for Higgsino detection \cite{2025arXiv250608084A}! Therefore, ALMA observations of M31* may bear the key role, complementing the findings of direct detection experiments by independent indirect revelation of Higgsino signal. 

\section{\label{sec:con}Conclusions and discussion}

M31* DM density spike was historically missed as a promising target for indirect WIMP searches. This work extends the study of M31* spike in gamma-ray band \cite{2026JCAP...04..002E} to the microwave band, where $e^\pm$ from annihilating DM produce the secondary synchrotron emission in strong MF of accretion flow. Estimation of this emission involved several major steps: modeling both DM density spike and the accretion flow medium, solving the transport equation for $e^\pm$, calculation of their synchrotron emission. Then I estimated ALMA sensitivity to WIMP parameters at the optimal frequency 100 GHz for various possible model parameter configurations. The following key conclusions can be drawn from the obtained results.
\begin{enumerate}
	\item The microwave band demonstrates significantly better sensitivity to WIMPs and narrower systematic uncertainty ranges in comparison with the gamma-ray band. Thus, it can be possible to reach the thermal WIMPs with $m_x\approx1$ TeV by reasonable ALMA exposures in the case of realistic spike density profile with $\gamma\approx1.75$; while CTA can probe in this case the annihilation cross sections, which are larger by 2-3 orders of magnitude!
	\item All thermal s-wave annihilating WIMP masses up to the unitarity limit $m_x\approx100$ TeV can be reached by ALMA only in the case of rather optimistic spike profile with $\gamma\approx2.0$ \emph{and} cuspy galactic DM halo density profile.
	\item The primary source of systematic model uncertainties is both density profiles. Uncertainties in SMBH mass, galactic central velocity dispersion, parameters of accretion flow medium have minor contributions in the total uncertainty budget. The latter is still very large and spans about 8 orders of magnitude in cross section at $m_x=1$ TeV for the considered parameter configurations.
	\item Higgsino signal might be visible for ALMA at 100 GHz with just $\sim10$ h exposure in an optimistic case scenario, although further robust identification of the signal source (i.e. distinguishing from the thermal background) may require much larger exposures.
	\item M31* accretion flow is expected to have lower MF strength, electron concentration and temperature in comparison with MW* flow. This enables two big advantages with respect to MW* case: the medium is much more transparent for the emission of interest (absorption becomes significant only at $\nu\lesssim40$ GHz) and generates much less thermal background/nuisance emission (the whole flow would be fainter by $\approx30$ times at 200 GHz at same distance).
	\item All other nearby SMBHs are expected to manifest significantly lower potential for WIMP detection at any wavelengths.
	\item I propose to conduct the deep dedicated observations of M31* by ALMA with 24--48 h exposures at each of 2--3 relevant frequencies, e.g. 50, 100, 200 GHz. Such observations will also yield useful data for accretion flow models as a by-product.
\end{enumerate}

Indeed, this work represents just an initial assessment and does/can not elaborate all the details of such complex physical system. I see the following potentially promising directions for further theoretical development.
\begin{enumerate}
	\item The detailed modeling of DM spike surface brightness distribution on the sky.
	\item Radiation transport modeling with absorption at low frequencies, which will allow to calibrate the whole model by existing M31* flux measurements at $\nu\leqslant20$ GHz, connect lower and higher frequency data, and further optimize the frequencies for WIMP signal search.
	\item Building the individual spike density model, which would take into account all the available information about M31 nucleus environment. So far a generic spike model was employed.
	\item Inclusion of the innermost region with $R\leqslant3R_\bullet$ into the model, where the effects of strong gravity may substantially enhance the emissivity \cite{2015ApJ...806..264S}.
	\item p-wave annihilating WIMPs might be considered.
\end{enumerate}

Summarizing, M31 is a fascinating object for DM indirect searches. Microwave observations of its SMBH vicinity may provide an outstanding chance to discover WIMPs. Indeed, next generation facilities, which are planned for future, will enhance this chance even more. These facilities include, in general, both: the microwave (e.g., ngVLA \cite{2018ASPC..517...15S}, ALMA--WSU \cite{2023pcsf.conf..304C}) and gamma-ray (e.g., SWGO \cite{SWGO}, AMS-100 \cite{AMS-100}, GAMMA-400 \cite{2017JPhCS.798a2011T}) band telescopes.

\appendix\section{\label{sec:a}Generalized solution of the transport equation with magnetization}

If to take into account the effect of magnetization of $e^\pm$ gas, which was mentioned in section \ref{sec:tr} and described by parameter $\zeta$, then the transport equation would have the form
\begin{equation*}
\frac{df}{dR}+\left(\frac{4\dot{p}_r}{pV_R}+\frac{3(1-\zeta)}{2R}\right)f = \frac{q_p}{V_R}
\end{equation*}
and the solution
\begin{align*}
f(R,p) &= \int\limits_{R_{sp}}^R\left(\frac{p_0}{p}\right)^4\left(\frac{R_0}{R}\right)^\frac{3+\zeta}{2}\frac{q_p(R_0,p_0)}{V_R(R_0)}dR_0, \\
p_0 &= \frac{p\sqrt{R_0}R^\lambda(3+2\zeta-\zeta^2)}{\sqrt{R}R_0^\lambda(3+2\zeta-\zeta^2)+2p(A(R^\lambda-R_0^\lambda)(3-\zeta)-A_\star(R^\lambda R_0^2-R_0^\lambda R^2)(1+\zeta))}, \\
\lambda &= (1+\zeta)/2;
\end{align*}
which is substantially heavier for numerical calculations in comparison with \eqref{eq:fsf},\eqref{eq:p0} with a little influence on the resulting emission flux.

\acknowledgments
I acknowledge the use of Wolfram Mathematica \circledR, ChatGPT \circledR~ (as advanced search engine) and WebPlotDigitizer \cite{WPD} software. I am also grateful to Maksim Lapunin for a friendly help with computer-related matters.

\bibliography{../../../universal}

\end{document}